\documentclass[%
reprint,
amsmath,amssymb,prl,
aps, groupedaddress,superscriptaddress
]{revtex4-1}

\usepackage{graphicx}
\usepackage{dcolumn}
\usepackage{bm}
\usepackage{braket}
\usepackage{mathtools}
\usepackage{bbm}
\usepackage{comment}
\usepackage{makecell}
\usepackage{afterpage}

\newcommand\blankpage{%
    \null
    \thispagestyle{empty}%
    \addtocounter{page}{-1}%
    \newpage}

\usepackage{hyperref}
\hypersetup{
	colorlinks=true,
	linkcolor=blue,
	citecolor=blue,
	filecolor=magenta,
	urlcolor=cyan
}

\def\iu{{\rm i}}
\def\hL{\mathcal{L}}

\def\hT{\mathcal{ T}}
\def\hQ{\mathcal{ Q}}
\def\hP{\mathcal{ P}}
\def\hE{\mathcal{ E}}

\def\hbetao{ \beta}

\def\halo{ \alpha}

\DeclareMathOperator{\tr}{Tr}
\DeclareMathOperator{\Tr}{Tr}

\graphicspath{{../}}

\begin{document}

\newcommand{\MM}[1]{[\textbf{MM} #1]}

\def\papertitle{{Kramers' degeneracy for open systems in thermal equilibrium}}

\def\tcm{TCM Group, Cavendish Laboratory, University of Cambridge, JJ Thomson Avenue, Cambridge, CB3 0HE, UK}
\def\ox{Rudolf Peierls Centre for Theoretical Physics, Clarendon Laboratory, Oxford OX1 3PU, UK}
\def\umd{Joint Quantum Institute, NIST/University of Maryland, College Park, Maryland 20742, USA}
\def\qcs{Joint Center for Quantum Information and Computer Science, NIST/University of Maryland, College Park, Maryland 20742 USA}
\def\fir{Department of Physics and Astronomy, University of Florence, Via G. Sansone 1, 50019 Sesto Fiorentino, Italy}

\title{\papertitle}
\author{Simon Lieu}
\affiliation{\umd}
\affiliation{\qcs}
\author{Max McGinley}
\affiliation{\ox}
\affiliation{\tcm}
\author{Oles Shtanko}
\affiliation{\umd}
\affiliation{\qcs}
\author{Nigel R.~Cooper}
\affiliation{\tcm}
\affiliation{\fir}
\author{Alexey V.~Gorshkov}
\affiliation{\umd}
\affiliation{\qcs}

\date{\today}

\begin{abstract}

Kramers' degeneracy theorem underpins many interesting effects in
quantum systems with time-reversal symmetry. We show that the generator of dynamics for Markovian open fermionic systems can exhibit an analogous degeneracy, protected by a combination of time-reversal symmetry \textit{and} the  microreversibility (detailed balance) property of systems at thermal equilibrium---the degeneracy is lifted if either condition is not met. We provide simple examples of this phenomenon  and show that the degeneracy is reflected in the  single-particle Green's functions. Furthermore, we show that certain experimental signatures of topological edge modes in open many-body systems can be protected by microreversibility in the same way. Our results highlight the importance of \textit{detailed balance} in characterizing open topological matter. 
\end{abstract}

\maketitle

Kramers' theorem is one of the oldest and most celebrated theorems in quantum mechanics. It states that  the Hamiltonian of a system with half-integer total spin will have a pairwise degenerate spectrum if time-reversal symmetry (TRS) is  preserved \cite{kramers1930, wigner1932}. This simple claim carries deep implications, ranging from the quantum spin Hall effect \cite{kane2005, bernevig2006} to the robustness of  superconductivity in disordered materials \cite{anderson1959}.

\begin{figure}
    \centering
    \includegraphics[scale=0.45]{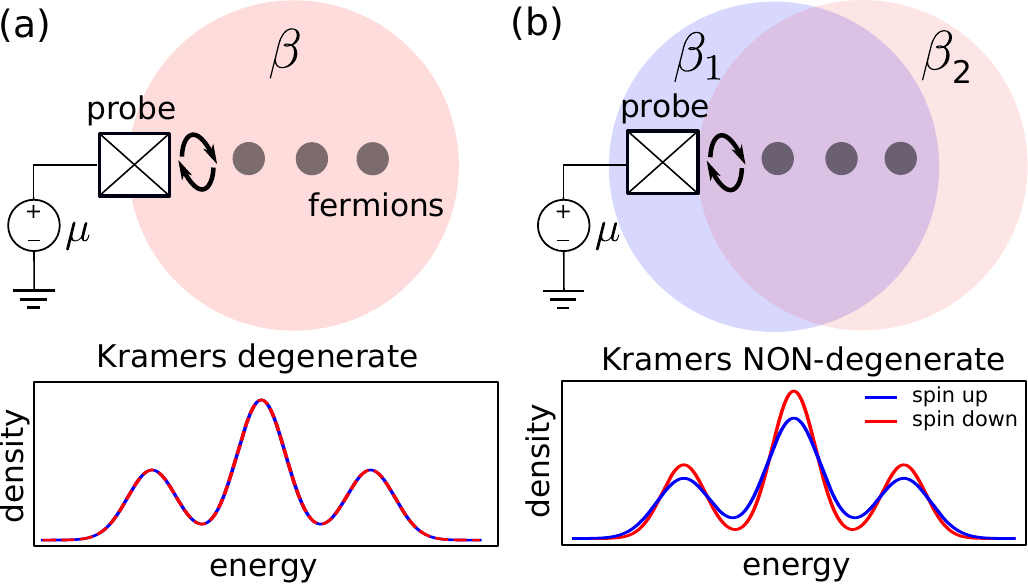}
    \caption{(a) A fermionic system (three small circles) coupled to an environment (big circle) in thermal equilibrium at temperature $\beta^{-1}$ has a Kramers' degeneracy. (b) The same system (three small circles) coupled to two baths (two big circles) at different temperatures $\beta_1^{-1}$, $\beta_2^{-1}$ will not be Kramers'-degenerate. This leads to a spin splitting of the Green's function \eqref{eq:retardedgreens}, which can be experimentally detected using, e.g.~spin-resolved tunneling spectroscopy (bottom panels). For both (a) and (b), the system-bath Hamiltonian obeys TRS: $[H_{SB},T]=0$. }
    \label{fig:bath}
\end{figure}

In this paper, we present an analogous degeneracy theorem that pertains to \textit{dissipative} many-body fermionic systems. Dissipative systems propagate irreversibly in time, i.e.~arbitrary initial states flow toward a (typically unique) steady state. Therefore it is not \textit{a priori} obvious that TRS should have any bearing on dynamics in this context. However, if the open system is at thermal equilibrium, then TRS manifests itself in terms of microreversibility (also known as quantum detailed balance) \cite{agarwal1973, majewski1984, alicki1976, chetrite2012, fagnola2008,jaksic2014, crooks2008, roberts2020, sieberer2015}, which has experimental consequences in the solid state \cite{benoit1986, jacquod2012, matthews2014, lopez2012}. 
Here we show that in Markovian open systems---which are governed by a Lindblad master equation with generator $\mathcal{L}$---the combination of TRS and microreversibility protect degeneracies in the spectrum of $\mathcal{L}$. An analogous non-equilibrium open system (e.g.~one coupled to two reservoirs held at different temperatures) will not exhibit this degeneracy; see Fig.~\ref{fig:bath}.

This degeneracy can be inferred in experiments that operate in a regime where coupling to an external environment cannot be ignored, such as noisy quantum simulators. Rather than being manifest in static properties of the steady state, the degeneracy of $\mathcal{L}$ can instead be inferred from time-dependent quantities describing the dynamical response about and/or approach to thermal equilibrium. As an example, we show that fermionic Green's functions and other related correlators \cite{bruus2004} show signatures of the degeneracy, making our results directly observable in the solid state \cite{oka2014spin} and atomic systems \cite{Dao2007}. 

Our theorem helps us identify an important connection between microreversibility and TRS-protected topological phases of matter. For closed systems, certain kinds of gapless edge modes can be attributed to Kramers' degeneracy  \cite{kane2005, bernevig2006, pollmann2010}. Although the irreversible effects of coupling such systems to an environment can spoil some of their TRS-protected properties \cite{mcginley2020}, here we demonstrate that other properties of the edge modes can persist in the open regime.

These results complement recent work regarding symmetries and topological phenomena in open systems \cite{diehl2011, lieu2020, moos2019, diehl2020, nori2020, wang2019, yoshida2020, chen2020, okuma2020, longhi2020, pan2020, sieberer2020, mcdonald2021, coser2019, deng2020}. Many of those studies are based on non-Hermitian Hamiltonians \cite{kawabata2019, nori2017, murakami2019, borgina2020, yao2018, malzard2015, loic2019,  lieu2018}, for which degeneracy theorems akin to Kramers' are known \cite{sato2012, zhou2019, kondo2020}, and have been central to explain symmetry-protected topological behavior. Our findings are distinct from such `non-Hermitian topological phenomena', which are primarily of relevance to photonic and classical mechanical systems \cite{bergholtz2019}. Indeed, the Lindblad master equation formalism used here is much more widely applicable to interacting, dissipative quantum matter.


\textit{Kramers' theorem for Hamiltonians.---}We review Kramers' theorem in closed fermionic systems. Since fermions have half-integer spin, time reversal is implemented by an antiunitary operator $T$ satisfying
\begin{align}
T^2=P, \qquad P=(-1)^{\hat{N}}, \qquad  [H,T]=0,
\label{eq:TRSOp}
\end{align}
where $\hat{N}$ is the total fermion number, and $P$ is henceforth referred to as `parity'. Note that $[P, T]=0$.  Physical Hamiltonians must conserve parity, $[H, P]= 0$, and so Fock space can be decomposed into even and odd parity sectors $\mathcal{H} = \mathcal{H}_+ \oplus \mathcal{H}_-$, wherein $H$ and $T$ are block diagonal
\begin{align}
H = \left(\begin{array}{cc}
H_+ & 0\\
0 & H_-
\end{array}\right),
\qquad 
T = \left(\begin{array}{cc}
U_+ & 0\\
0 & U_-
\end{array}\right) K.
\label{eq:blockdiaghamiltonian}
\end{align}
Here, $K$ takes the complex conjugate of any scalar to its right, and   $U_\pm^{\vphantom{*}} U_\pm^*=\pm\mathbb{I}$.

Kramers' theorem states that for a TRS-invariant Hamiltonian (i.e.\ when $T H T^{-1} = H$), the eigenvalues of $H_-$ must be twofold degenerate. Specifically, any eigenstate $\ket{\psi_-}$ with odd parity has a time-reversed partner $T\ket{\psi_-}$ which is also an eigenstate of the same energy and is orthogonal to $\ket{\psi_-}$. In contrast, $H_+$ is generically non-degenerate.

\textit{Symmetries of open quantum systems.---}For the purposes of this Letter, we will focus on Markovian open systems described by a density matrix $\rho$, whose dynamics is governed by a Lindblad master equation  \cite{lindblad1976, gorini1976}
\begin{align}
    \frac{d \rho}{dt} =  \mathcal{L}(\rho) = -i [H, \rho] +  \sum_i \left( 2 L_i \rho L_i^\dagger - \{L^\dagger_i L_i, \rho \}\right).
    \label{eq:lindbladeq}
\end{align}
Here, the Hamiltonian part $H$ describes the coherent evolution of the system, and the dissipators $L_{i}$ arise from coupling to an environment. 
The generating superoperator $\mathcal{L}$ is often referred to as the `Lindbladian'.


We define time-reversal and parity superoperators ($\mathcal{ T}$ and $\mathcal{ P}$, respectively) which specify how these symmetries transform the state $\rho(t)$. These act as $\mathcal{ T}[\rho] = T \rho T^{-1}$ and $\mathcal{ P}[\rho] = P \rho P^{-1}$. The space of operators can be split into even and odd superparity sectors $\mathcal{B}(\mathcal{H}) = \mathcal{B}(\mathcal{H})_+ \oplus \mathcal{B}(\mathcal{H})_-$, where $\mathcal{B}(\mathcal{H})_\pm$ contains operators satisfying $\mathcal{ P}(A) = \pm A$ (where $A$ is an arbitrary operator). Specifically, $\mathcal{B}(\mathcal{H})_+$ contains operators of the form $|\text{even}\rangle \langle \text{even}|$ or $|\text{odd}\rangle \langle \text{odd}|$, and operators in $\mathcal{B}(\mathcal{H})_-$ are $|\text{even}\rangle \langle \text{odd}|$ or $|\text{odd}\rangle \langle \text{even}|$, which are traceless. (We use the convention $P |\text{even}\rangle = +|\text{even}\rangle, P |\text{odd}\rangle = -  |\text{odd}\rangle$.) Since coherent superpositions of states with opposite fermion parity are not physical,
the system density matrix must belong to $\mathcal{B}(\mathcal{H})_+$ (note that this does not prohibit classical mixtures of wavefunctions with opposite parity). Accordingly, physical generators must satisfy $[\mathcal{ L}, \mathcal{ P}] = 0$. By analogy to Eq.~\eqref{eq:blockdiaghamiltonian}, the matrix representations of $\mathcal{ L}$ and $\mathcal{ T}$ then become block diagonal:
\begin{align} \label{eq:blockdiaglindblad}
\mathcal{L} = \left(\begin{array}{cc}
\mathcal{L}_+ & 0\\
0 & \mathcal{L}_-
\end{array}\right),
\qquad 
\mathcal{T} = \left(\begin{array}{cc}
\mathcal{U}_+ & 0\\
0 & \mathcal{U}_-
\end{array}\right) K.
\end{align}
From Eq.~\eqref{eq:TRSOp}, we have $\mathcal{ T}^2 = \mathcal{ P}$, and hence $ \mathcal{U}_\pm^{\vphantom{*}} \mathcal{U}_\pm^*=\pm\mathbb{I}$. Evidently, the superoperator $\mathcal{ T}^2$ 
leaves operators in $\mathcal{B}(\mathcal{H})_-$ invariant only up to a $(-1)$ phase.  This contrasts with systems made up of bosonic or spin degrees of freedom, where $T^2 = \pm 1$, and hence $\hT^2 = \mathbb{I}$, even for half-integer spins.

Since $\rho(t)$ belongs to $\mathcal{B}(\mathcal{H})_+$, it is often stated that the odd-superparity part of the Lindbladian $\mathcal{L}_-$ is unphysical. However, this is not true if we ask about the joint state of two fermionic systems. If system $S$ and some probe $R$ evolve independently under Lindbladians $\mathcal{L}_{S,R}$, then their joint state $\rho_{SR}(t)$ evolves as $\partial_t \rho_{SR} = (\mathcal{L}_S \otimes \text{id}_R + \text{id}_S \otimes \mathcal{L}_R)[\rho_{SR}]$. When the two are coupled, the overall superparity $\mathcal{P}_{\rm tot} = \mathcal{P}_S \mathcal{P}_R$ must still be even ($\mathcal{P}_S$ and $\mathcal{P}_R$ are superparities for $S$ and $R$, respectively), however the state $\rho_{SR}(t)$ may contain components which are odd under $\mathcal{P}_S$ and $\mathcal{P}_R$ separately, e.g.~if a fermion is in a superposition between $S$ and $R$. These components will evolve under $\mathcal{L}_{S,-}$ and $\mathcal{L}_{R,-}$. We will later identify specific physical observables that can be used to infer properties of $\mathcal{L}_-$. We  now establish a degeneracy theorem analogous to Kramers' which applies to the spectrum of $\mathcal{L}_-$.



\textit{Kramers' degeneracy for Lindbladians.}---Because open quantum systems evolve irreversibly in time, TRS cannot be expected to play the same role as in closed systems. Indeed, if the Hamiltonian $H$ is TRS-invariant in the sense of Eq.~\eqref{eq:TRSOp}, then we have $\mathcal{ T}^{-1} \mathcal{ L} \mathcal{ T} = -\mathcal{ L}$. This relation is incompatible with a non-trivial dissipative part of Eq.~\eqref{eq:lindbladeq}, which is negative semi-definite \cite{breuer2002}. One way in which an antiunitary symmetry
can be imposed on open systems without such an inconsistency is to demand that the Hamiltonian be \textit{odd} under TRS, i.e.~ $T H T^{-1} = -H$, in which case $\mathcal{ T}^{-1} \mathcal{ L} \mathcal{ T} = \mathcal{L}$ can be satisfied \cite{altland2020}. Although such a symmetry is mathematically well-defined, it does not physically correspond to time-reversal in the closed system limit.

Instead, for systems in thermal equilibrium, time reversal symmetry of the system and environment degrees of freedom naturally gives rise to a  ``microreversibility'' property, otherwise known as detailed balance. This condition relates the rate of each possible physical process to the rate of its time-reversed process. Mathematically, a Lindbladian that respects detailed balance satisfies a superoperator equation \cite{agarwal1973,majewski1984, chetrite2012}
\begin{align} \label{eq:micro}
\hL^\dagger =   \hQ^{-1}  \hT ^{-1} \hL  \hT  \hQ,
\end{align}
where $\hQ$ acts as $\hQ[A] = q A$, and  $q = \exp{ [-\beta H]} / Z$ is the density matrix in the Gibbs ensemble at inverse temperature $\beta$ with respect to the system Hamiltonian $H$ ($Z$ is the partition function).  We have defined
\begin{align}  \label{eq:adjoint}
\mathcal{ L}^\dagger [\rho] = +i [H, \rho] +  \sum_i \left( 2  L_i^\dagger \rho L_i - \{ L^\dagger_i  L_i,\rho \}\right),
\end{align}
which is the usual adjoint of $\mathcal{L}$ with respect to the Hilbert-Schmidt inner product $\braket{A, B} \coloneqq \Tr[A^\dagger B]/ \Tr \mathbb{I}$.

 A simple example of jump operators satisfying \eqref{eq:micro} is a pair $L_1 = \sqrt{\gamma_1} V$, $L_2 = \sqrt{\gamma_2} V^\dagger$, where $V$ is TRS-invariant up to a phase, $\hT[V] = e^{\iu \theta} V$ \footnote{The phase $\theta$ carry no physical significance, since jump operators can be changed by a transform $L_i \rightarrow e^{ i  \chi}L_i$ without changing the Lindbladian \eqref{eq:lindbladeq}.}, and acts as a lowering operator $[H,V] = -\omega V$. The temperature is implicitly determined by $\gamma_1/\gamma_2 = e^{\beta \omega}$. In words, $L_1$ lowers the system energy by $\omega$ at a given rate, and $L_2$ does the opposite at a rate that differs by a Boltzman factor. This is the essence of the detailed balance condition. More generally, Eq.~\eqref{eq:micro} is naturally satisfied when a time-reversal symmetric system is coupled to TRS-respecting environment at thermal equilibrium with temperature $\beta^{-1}$ \cite{SM}.
 

We now state the main claim of this work: Suppose a fermionic Lindbladian satisfies  microreversibility via Eq.~\eqref{eq:micro}; then  the odd-superparity superoperator $\mathcal{L}_-$ [see Eq.\ \eqref{eq:blockdiaglindblad}] is guaranteed to have a twofold degenerate spectrum. Our proof proceeds as follows. The odd parity part of the Lindbladian satisfies
\begin{align}
\hL^\dagger_- =   \hQ^{-1}_-  \hT_- ^{-1} \hL_-  \hT_-  \hQ_-,
\end{align}
where $\mathcal{T}^2_-=-\mathbb{I}, \mathcal{T}_- = \mathcal{U}_- K$. We drop the ``$-$''  index. Define  right and left eigenoperators
\begin{align}
\hL ( r_i ) =  \Lambda_i  r_i  , \qquad \hL^\dagger (l_i)  =  \Lambda_i^* l_i ,
\end{align}
with $ \Tr [ l_i^\dagger  r_j ] = \delta_{ij}$. Substituting the expression for microreversibility into the left eigenoperator  equation leads to 
\begin{align}
\hQ^{-1}  \hT^{-1}  \hL  \hT  \hQ (l_i)  = \Lambda_i^*  l_i  \Rightarrow  \hL  \hT  \hQ ( l_i ) = \Lambda_i \hT  \hQ (l_i).
\end{align}
We find that  $r_i  $ and $\hT  \hQ  (l_i)  $ are both right eigenoperators of $\hL$ with eigenvalue $ \Lambda_i$.  However, one can show that $\Tr[  l_i^\dagger \hT  \hQ ( l_i)] =0$ (see the Supplemental Material \cite{SM}). Since $\Tr[l_i^\dagger r_i ] =1$, we find that  $ r_i  $ and $\hT  \hQ ( l_i)$ are linearly independent  eigenoperators, and hence the complex Lindblad spectrum in the odd superparity sector must be twofold degenerate.

In contrast to the  Kramers’ theorem for closed systems, our result explicitly relies on the presence of thermal equilibrium. Intuition can be gained from microscopic considerations: Linear response about a thermal  state can be formulated in terms of the (Kramers’ degenerate) eigenstates of the system-bath Hamiltonian. A system’s response in thermal equilibrium should thus be sensitive to the TRS of the microscopic Hamiltonian. Our work captures this behavior from the perspective of the system's master equation.

Our analysis also suggests that a similar Kramers’ degeneracy is present in the spectrum of a thermal quantum channel superoperator, which can describe the evolution of a system coupled to a non-Markovian bath \cite{SM}. 

\begin{figure}
    \centering
    \includegraphics[scale=0.45]{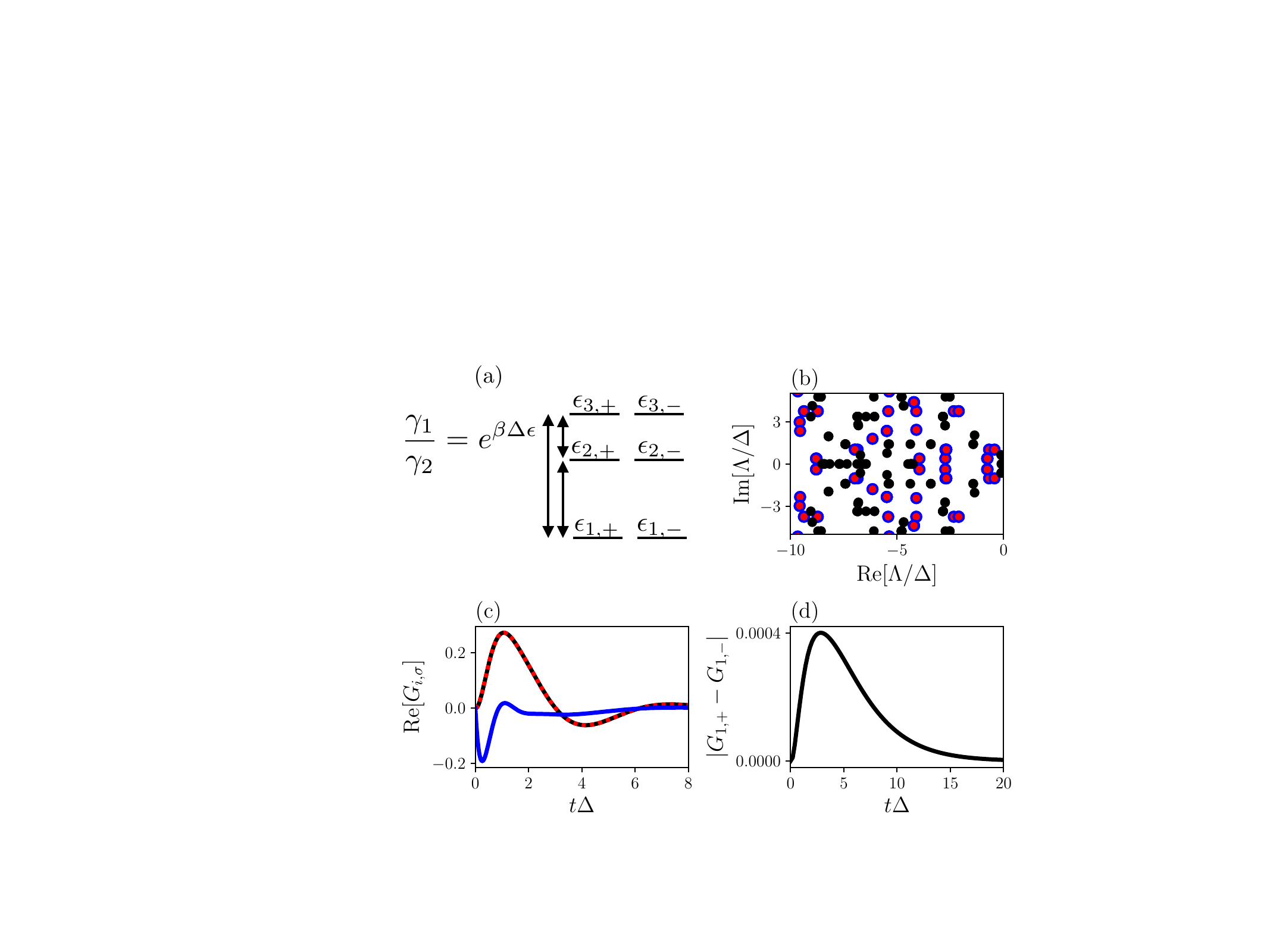}
    \caption{(a) Three doubly degenerate energy levels ($N=3$) labeled by $\epsilon_{i, \pm}$. TRS-respecting dissipators cause jumps between energy levels at a rate consistent with detailed balance in thermal equilibrium at temperature $\beta^{-1}$. (b)  Lindblad spectrum for  the model described in the main text (for states with 0, 1, or 2 occupied fermions) at a temperature $\beta \Delta=1$, with $N=3$,  $\Delta_d/\Delta = 0.1$, and $g/\Delta=1$. The odd-superparity sector (blue and red dots) is twofold degenerate, while the even sector (black dots) is not. (c) Real part of the retarded Green's function for same parameters as in (b). The Kramers' degeneracy ensures that $G_{i, \sigma} = G_{i,-\sigma}$ in the expectation value of the thermal steady state with one occupied fermionic mode: black line and red dashed line correspond to $\textrm{Re}[G_{1,+}]$ and  $\textrm{Re}[G_{1,-}]$, respectively. In general, Green's function pairs other than $G_{i, \pm}$ are non-degenerate: blue line corresponds to $\textrm{Re}[G_{2,+}]$.  (d) A system that is coupled to \textit{two} thermal baths $g_1/\Delta=1, g_2/\Delta=0.4$ at different temperatures: $\beta_1 \Delta = 1, \beta_2 \Delta = 10$ [otherwise same parameters as in (b)]. Microreversibility is broken, hence the Green's functions for $i,+$ and $i,-$ split.}
    \label{fig:spec}
\end{figure}

\textit{Example: random quadratic Hamiltonian.}---We confirm the generalized Kramers' theorem via an example. Consider a system of spin-1/2 fermions with $N$ two-fold degenerate single-particle orbitals. The most general particle-conserving quadratic Hamiltonian is
\begin{align} \label{eq:hamex}
H = \sum_{ij; \sigma, \sigma'}  H_{ij,\sigma\sigma'} f_{i, \sigma}^\dagger f_{j,\sigma'} = \sum_{k=1}^N  \sum_{\tau = \pm}  \epsilon_k  c_{k, \tau}^\dagger c_{k,\tau}
\end{align}
with $i,j \in [1,\ldots, N]$ and $\sigma, \sigma' \in \pm$. We impose a TRS on the Hamiltonian: $[H, T]=0$, such that the single-particle spectrum is twofold degenerate, with $T f_{i, \sigma}  T^{-1} = \sigma f_{i, -\sigma}$, $T c_{k, \tau} T^{-1} =  \tau  c_{k, -\tau}$.  Let us define the following dissipators
\begin{align}  \label{eq:disex}
L_{1, pq} &= \sqrt{\gamma_{1, pq}} \sum_{\tau = \pm} \left( c_{q, \tau}^\dagger c_{p,\tau} + \tau c_{q, \tau}^\dagger c_{p,-\tau}  \right) , \\
L_{2, pq} &=  \sqrt{\gamma_{2, pq}} \sum_{\tau = \pm} \left( c_{p,\tau}^\dagger c_{q, \tau}  + \tau  c_{p,-\tau}^\dagger  c_{q, \tau} \right) , \label{eq:disex2}
\end{align}
for $\gamma_{1,pq} = g [n_\beta(\epsilon_p - \epsilon_q) +1]$, $ \gamma_{2,pq} =g n_\beta(\epsilon_p - \epsilon_q) $, $n_\beta(\omega) = (e^{\beta \omega}-1)^{-1}$ [the Bose function], and   $\epsilon_p  > \epsilon_q$. Physically, such terms can appear due to coupling to a bosonic bath (since the dissipators are quadratic in fermions). $L_{1}$ represents a  process that lowers the energy of the system;  $L_{2}$ raises the energy. The $L$s  satisfy:  $[H, L_{1,pq}] = (\epsilon_q - \epsilon_p) L_{1,pq}, L_{2,pq} \sim L_{1,pq}^\dagger, \gamma_{1,pq} / \gamma_{2,pq} = e^{\beta (\epsilon_{p} - \epsilon_q)}$, and $ [ L_{1/2,pq}, T] = 0 $,   thus respecting microreversibility  [see Fig.~\ref{fig:spec}(a)]. We include these dissipators  between each pair of energy levels in the system, and also consider  uniform dephasing in the energy basis: $L_{d, i, \pm} = \sqrt{\Delta_d} c_{i,\pm}^\dagger c_{i,\pm}$ (preserves microreversibility).  For a fixed number of fermions in the system, the thermal  state is the unique steady state. 

Fig.~\ref{fig:spec}(b) plots the Lindblad spectrum associated with a random TRS-respecting Hamiltonian ($|H_{ij}|  \in [0, \Delta]$; $\Delta$ sets the energy scale) coupled to a thermal bath for states with up to two occupied fermionic modes. The spectrum of the odd superparity sector (blue and red dots) is indeed twofold degenerate, whilst the spectrum of the even sector (black dots) is not. Note that the Kramers' theorem does not imply a degeneracy of the steady state.

How can an open system violate microreversibility (and hence break its degeneracy)?
One obvious way involves a system-environment coupling that directly violates TRS.  However, a more subtle way to break degeneracy involves coupling the system to a non-equilibrium environment. Consider a system that is connected to  \textit{two} thermal baths, each at a different temperature. Even if all system-environment couplings respect TRS, the system will host a \textit{non-equilibrium} (non-thermal) steady state, and hence microreversibility will be violated. We have numerically verified that the degeneracy of $\mathcal{L}_-$ for the spin-$1/2$ system is broken when coupled to baths at different temperatures. [We note that such non-equilibrium splitting requires quartic terms (quadratic dissipators) in the master equation (see SM \cite{SM})].

\textit{Physical observables.---}As argued above, $\mathcal{L}_-$ only governs dynamics in scenarios where fermions can move between the system ($S$) and some external `probe' ($R$). Therefore, rather than looking at expectation values of the system density matrix $\Tr[O_S \rho_S(t)]$ (which are determined by $\mathcal{L}_+$), we suggest that the Kramers'-like degeneracy of $\mathcal{L}_-$ can be detected in the retarded Green's function of the steady state $\rho_{\rm SS}$:
\begin{align}
G_{i,\sigma}^R(t) = -i \Theta(t) \text{Tr}[\{ f_{i,\sigma}(t), f_{i,\sigma}^\dagger \} \rho_{\rm SS}].
\label{eq:retardedgreens}
\end{align}
Here, $\Theta(t)$ is the Heaviside step function, and we work in the Heisenberg picture where operators evolve as $A(t) = e^{\mathcal{L}^\dagger t}[A]$ (although this expression should be slightly modified for open systems coupled to fermionic baths; see the Supplemental Material \cite{SM}). 

The Green's function \eqref{eq:retardedgreens} can be measured in solid state systems using, e.g.~photoemission or tunnelling spectroscopy, which indeed involve fermions moving in/out of the system \cite{oka2014spin}. Probing single-particle Green's functions  in ultracold atoms is more challenging, but protocols involving stimulated Raman spectroscopy have been developed \cite{Dao2007}.
More concretely, $G_{i,\sigma}^R(t)$ is sensitive to the Kramers' degeneracy of $\mathcal{L}_-$ because $f_{i,\sigma} \in \mathcal{B}(\mathcal{H})_-$ is superparity-odd, and so the time evolution of \eqref{eq:retardedgreens} is governed by $\mathcal{L}_-$.  
The generalized Kramers' theorem ensures the relation:  $G_{i,\sigma}^R(t) = G_{i,-\sigma}^R(t)$ \cite{SM}, which we confirm numerically in Fig.~\ref{fig:spec}(c). However, when the system is coupled to two baths at different temperatures, microreversibility is broken and the Green's functions differ for opposite spins [Fig.~\ref{fig:spec}(d)].  We note that the Fourier transform of the temporal Green's function is directly probed in solid-state electron-tunneling experiments (see below, and  the SM \cite{SM})

It is rather natural that microreversibility has implications for response functions such as \eqref{eq:retardedgreens}; indeed, the very definition of microreversibility is sometimes framed in terms of fluctuation-dissipation relations for steady-state correlators \cite{agarwal1973,  lopez2012}. The above demonstrates that the correlation functions of superparity-odd operators have a particular structure associated with the Kramers' degeneracy of $\mathcal{L}_-$.

\textit{Degenerate zero-bias peak from microreversibility.---}Kramers' degeneracy plays an important role in determining the stability of symmetry-protected topological edge modes of Hamiltonians \cite{kane2005, ryu2010}, e.g.~the presence of spinful TRS ensures that   a pair of Majorana zero modes in 1D cannot gap. For closed systems, degenerate Majorana modes are detectable via degenerate spin-resolved  tunneling spectroscopy at the edge of the superconductor \cite{bruus2004}. Here, we show that the spin-resolved tunneling spectra remain unspilt if the superconductor is coupled to a thermal bath with TRS-respecting terms, while a splitting can arise if microreversibility is violated.

Consider the following  spin-$1/2$ Kitaev chain
\begin{align} \label{eq:diii}
H = &-i \sum_{j=1, \sigma=\pm}^{j=N}\left[  u_1 a_{ j, \sigma} b_{j, \sigma}+ u_2 a_{j, \sigma} b_{ j, -\sigma}  \right]  \\
 &-i \sum_{j=1, \sigma=\pm}^{j=N-1}  \left[ v_1 b_{ j, \sigma} a_{ j+1, \sigma}  + v_2 b_{ j, \sigma} a_{ j+1, -\sigma}  \right]  \nonumber,
\end{align}
where $a_{j,\sigma},b_{j,\sigma}$ are Majoranas corresponding to site $j$ with spin $\sigma$. This model can be diagonalized as: $ H = \sum_{j=1}^N \sum_{\tau =\pm} \epsilon_j d_{j, \tau}^\dagger d_{j,\tau}  $,  where $ T d_{j, \tau}  T^{-1} = \tau d_{j,-\tau}$.  We include the same thermal dissipators as before [see Eq.~\eqref{eq:disex}] between each pair of energy levels $p,q$  in the system at inverse temperature $\beta$. While such dissipators are manifestly non-local, they can arise from local system-bath coupling \cite{breuer2002}.  The weak-coupling Markovian approximation is commonly used in studying topological matter connected to a thermal bath \cite{alicki2009, bravyi2013}. To break microreversibility, we  consider a non-equilibrium bath that removes pairs of fermions on each site: $L_j = \sqrt{g_{\text{neq}}} \psi_{j, +} \psi_{j, -}$ where we define the complex fermion: $\psi_{j, \sigma} = a_{j, \sigma} + i b_{j, \sigma} $. These dissipators  obey $[L_j,  T]=0$ but do not evolve the system toward a thermal state.

\begin{figure}
    \centering
    \includegraphics[scale=0.27]{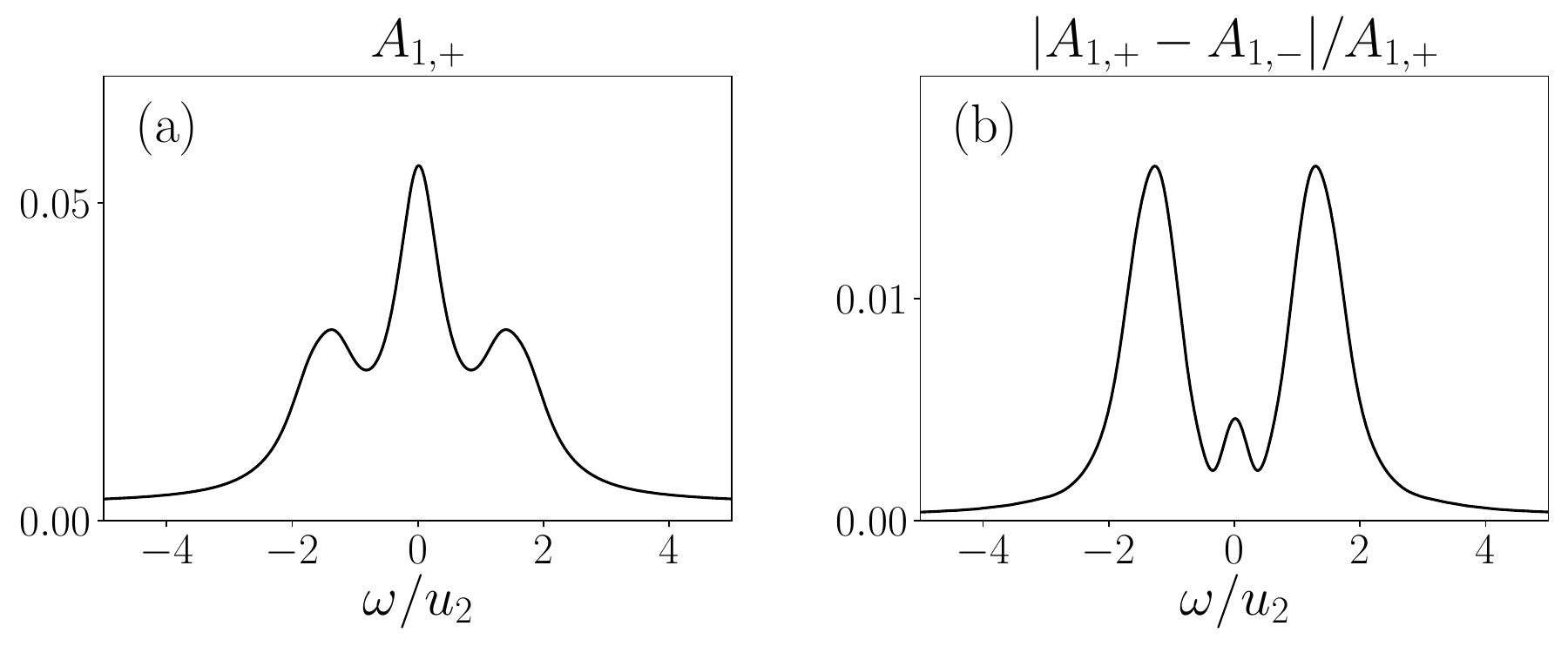}
    \caption{ Steady-state, spin-resolved  spectral function at the boundary of the Majorana chain \eqref{eq:diii} coupled to two baths found via exact diagonalization. (a) Zero-bias peak due to the Majorana modes. The width of the peaks is set by the temperature. (b) Splitting between spin-up and spin-down spectral functions due to breaking of microreversibility via a non-equilibrium environment. Parameters: $N=3, u_1 / v_1 = 0.25 , u_2/v_1 = 0.25 , v_2 / v_1 =  0.5, \beta v_1 = 3.3, g / v_1 = 0.05 , g_{\text{neq}} / v_1 = 2$.  }
    \label{fig:diii}
\end{figure}

The spectral degeneracy (or its splitting) can be experimentally detected using spin-resolved tunneling spectroscopy \cite{oka2014spin} (see Fig.~\ref{fig:bath}). For a zero-temperature probe, the current-voltage characteristic of the tunnel junction is $\partial I_\sigma/\partial \mu \propto A_{1,\sigma}(-\mu)$, where  $A_{i,\sigma}(\omega) = -\text{Im}[G_{i, \sigma} ^R(\omega)]$ is the spectral function, $I_\sigma$ is spin-$\sigma$ current and $\mu$ is the chemical potential (see SM \cite{SM}). Fig.~\ref{fig:diii} plots the spin-up spectral function for the probe attached to the boundary, and the relative difference between the spin-up and spin-down spectral functions for the chain coupled to two baths. The zero-bias peak in the topological phase corresponds to the boundary-localized Majoranas. The splitting that emerges between the spin up and down spectral functions is due to the non-equilibrium setup.
This splitting vanishes if the non-equilibrium bath is turned off and micoreversibility is restored (not shown). While isolating a Majorana chain from its larger environment is a  difficult task, our result suggests that 
signatures of the TRS-protected Majorana modes should still be observable in experiment provided that the environment is \textit{thermal}.

\textit{Conclusions and outlook.}---We have proved a generalization of  Kramers'  theorem for open quantum systems, and shown that it  has implications for symmetry-protected topological phases. 


Future work should further investigate the role that detailed balance plays in protecting topological signatures and phases of open quantum systems. For example, it is known that non-Hermitian generalizations of time-reversal symmetry can prevent the non-Hermitian skin effect \cite{kawabata2019}, i.e.~extreme spectral sensitivity to boundary conditions \cite{yao2018, kunst2018}. Can  microreversibility guarantee the absence of the skin effect in Lindbladians? 

Imposing TRS on a Hamiltonian allows us to identify topologically-distinct ground states which cannot smoothly evolve into one another without closing the energy gap \cite{pollmann2010}. It is currently unclear whether
microreversibility has similar implications for the topological properties of the steady-state density matrix  \cite{coser2019}.

 
Finally, we note that microreversibility may also have interesting implications for the random-matrix theory of open systems  \cite{denisov2019, can2019, prosen2020, wang2020}, e.g.~the non-Hermitian random matrix generalization of class AII   might constrain the spectral statistics of the odd-superparity sector of microreversible Lindbladians \cite{hamazaki2020}. 
 
\begin{acknowledgments}
S.L.\ was supported by the NIST NRC Research Postdoctoral Associateship Award. M.M.\ was supported by an EPSRC studentship. O.S.\ and A.V.G.\ acknowledge funding by AFOSR MURI, NSF PFCQC program, AFOSR, U.S.\ Department of Energy Award No.\ DE-SC0019449, DoE ASCR Quantum Testbed Pathfinder program (award No.\ DE-SC0019040), DoE ASCR Accelerated Research in Quantum Computing program (award No.\ DE-SC0020312), and ARO MURI. N.R.C.\ was supported by EPSRC Grants  EP/P034616/1 and EP/P009565/1 and by an Investigator Award of the Simons Foundation.
\end{acknowledgments}

\bibliography{kram.bib}

\begin{thebibliography}{68}%
\makeatletter
\providecommand \@ifxundefined [1]{%
 \@ifx{#1\undefined}
}%
\providecommand \@ifnum [1]{%
 \ifnum #1\expandafter \@firstoftwo
 \else \expandafter \@secondoftwo
 \fi
}%
\providecommand \@ifx [1]{%
 \ifx #1\expandafter \@firstoftwo
 \else \expandafter \@secondoftwo
 \fi
}%
\providecommand \natexlab [1]{#1}%
\providecommand \enquote  [1]{``#1''}%
\providecommand \bibnamefont  [1]{#1}%
\providecommand \bibfnamefont [1]{#1}%
\providecommand \citenamefont [1]{#1}%
\providecommand \href@noop [0]{\@secondoftwo}%
\providecommand \href [0]{\begingroup \@sanitize@url \@href}%
\providecommand \@href[1]{\@@startlink{#1}\@@href}%
\providecommand \@@href[1]{\endgroup#1\@@endlink}%
\providecommand \@sanitize@url [0]{\catcode `\\12\catcode `\$12\catcode
  `\&12\catcode `\#12\catcode `\^12\catcode `\_12\catcode `\%12\relax}%
\providecommand \@@startlink[1]{}%
\providecommand \@@endlink[0]{}%
\providecommand \url  [0]{\begingroup\@sanitize@url \@url }%
\providecommand \@url [1]{\endgroup\@href {#1}{\urlprefix }}%
\providecommand \urlprefix  [0]{URL }%
\providecommand \Eprint [0]{\href }%
\providecommand \doibase [0]{http://dx.doi.org/}%
\providecommand \selectlanguage [0]{\@gobble}%
\providecommand \bibinfo  [0]{\@secondoftwo}%
\providecommand \bibfield  [0]{\@secondoftwo}%
\providecommand \translation [1]{[#1]}%
\providecommand \BibitemOpen [0]{}%
\providecommand \bibitemStop [0]{}%
\providecommand \bibitemNoStop [0]{.\EOS\space}%
\providecommand \EOS [0]{\spacefactor3000\relax}%
\providecommand \BibitemShut  [1]{\csname bibitem#1\endcsname}%
\let\auto@bib@innerbib\@empty
\bibitem [{\citenamefont {Kramers}(1930)}]{kramers1930}%
  \BibitemOpen
  \bibfield  {author} {\bibinfo {author} {\bibfnamefont {H.~A.}\ \bibnamefont
  {Kramers}},\ }\href@noop {} {\bibfield  {journal} {\bibinfo  {journal}
  {Koninkl. Ned. Akad. Wetenschap., Proc.}\ }\textbf {\bibinfo {volume} {33}},\
  \bibinfo {pages} {959} (\bibinfo {year} {1930})}\BibitemShut {NoStop}%
\bibitem [{\citenamefont {Wigner}(1932)}]{wigner1932}%
  \BibitemOpen
  \bibfield  {author} {\bibinfo {author} {\bibfnamefont {E.}~\bibnamefont
  {Wigner}},\ }\href@noop {} {\bibfield  {journal} {\bibinfo  {journal}
  {Nachrichten von der Gesellschaft der Wissenschaften zu G{\"o}ttingen,
  Mathematisch-Physikalische Klasse}\ }\textbf {\bibinfo {volume} {1932}},\
  \bibinfo {pages} {546} (\bibinfo {year} {1932})}\BibitemShut {NoStop}%
\bibitem [{\citenamefont {Kane}\ and\ \citenamefont {Mele}(2005)}]{kane2005}%
  \BibitemOpen
  \bibfield  {author} {\bibinfo {author} {\bibfnamefont {C.~L.}\ \bibnamefont
  {Kane}}\ and\ \bibinfo {author} {\bibfnamefont {E.~J.}\ \bibnamefont
  {Mele}},\ }\href {\doibase 10.1103/PhysRevLett.95.226801} {\bibfield
  {journal} {\bibinfo  {journal} {Phys. Rev. Lett.}\ }\textbf {\bibinfo
  {volume} {95}},\ \bibinfo {pages} {226801} (\bibinfo {year}
  {2005})}\BibitemShut {NoStop}%
\bibitem [{\citenamefont {Bernevig}\ and\ \citenamefont
  {Zhang}(2006)}]{bernevig2006}%
  \BibitemOpen
  \bibfield  {author} {\bibinfo {author} {\bibfnamefont {B.~A.}\ \bibnamefont
  {Bernevig}}\ and\ \bibinfo {author} {\bibfnamefont {S.-C.}\ \bibnamefont
  {Zhang}},\ }\href {\doibase 10.1103/PhysRevLett.96.106802} {\bibfield
  {journal} {\bibinfo  {journal} {Phys. Rev. Lett.}\ }\textbf {\bibinfo
  {volume} {96}},\ \bibinfo {pages} {106802} (\bibinfo {year}
  {2006})}\BibitemShut {NoStop}%
\bibitem [{\citenamefont {Anderson}(1959)}]{anderson1959}%
  \BibitemOpen
  \bibfield  {author} {\bibinfo {author} {\bibfnamefont {P.}~\bibnamefont
  {Anderson}},\ }\href {\doibase https://doi.org/10.1016/0022-3697(59)90036-8}
  {\bibfield  {journal} {\bibinfo  {journal} {J. Phys. Chem. Solids}\ }\textbf
  {\bibinfo {volume} {11}},\ \bibinfo {pages} {26 } (\bibinfo {year}
  {1959})}\BibitemShut {NoStop}%
\bibitem [{\citenamefont {Agarwal}(1973)}]{agarwal1973}%
  \BibitemOpen
  \bibfield  {author} {\bibinfo {author} {\bibfnamefont {G.}~\bibnamefont
  {Agarwal}},\ }\href@noop {} {\bibfield  {journal} {\bibinfo  {journal} {Z.
  Phys. A}\ }\textbf {\bibinfo {volume} {258}},\ \bibinfo {pages} {409}
  (\bibinfo {year} {1973})}\BibitemShut {NoStop}%
\bibitem [{\citenamefont {Majewski}(1984)}]{majewski1984}%
  \BibitemOpen
  \bibfield  {author} {\bibinfo {author} {\bibfnamefont {W.~A.}\ \bibnamefont
  {Majewski}},\ }\href {\doibase 10.1063/1.526164} {\bibfield  {journal}
  {\bibinfo  {journal} {J. Math. Phys.}\ }\textbf {\bibinfo {volume} {25}},\
  \bibinfo {pages} {614} (\bibinfo {year} {1984})}\BibitemShut {NoStop}%
\bibitem [{\citenamefont {Alicki}(1976)}]{alicki1976}%
  \BibitemOpen
  \bibfield  {author} {\bibinfo {author} {\bibfnamefont {R.}~\bibnamefont
  {Alicki}},\ }\href {\doibase https://doi.org/10.1016/0034-4877(76)90046-X}
  {\bibfield  {journal} {\bibinfo  {journal} {Rep. Math. Phys}\ }\textbf
  {\bibinfo {volume} {10}},\ \bibinfo {pages} {249 } (\bibinfo {year}
  {1976})}\BibitemShut {NoStop}%
\bibitem [{\citenamefont {Chetrite}\ and\ \citenamefont
  {Mallick}(2012)}]{chetrite2012}%
  \BibitemOpen
  \bibfield  {author} {\bibinfo {author} {\bibfnamefont {R.}~\bibnamefont
  {Chetrite}}\ and\ \bibinfo {author} {\bibfnamefont {K.}~\bibnamefont
  {Mallick}},\ }\href {https://doi.org/10.1007/s10955-012-0557-z} {\bibfield
  {journal} {\bibinfo  {journal} {J. Stat. Phys.}\ }\textbf {\bibinfo {volume}
  {148}},\ \bibinfo {pages} {480} (\bibinfo {year} {2012})}\BibitemShut
  {NoStop}%
\bibitem [{\citenamefont {Fagnola}\ and\ \citenamefont
  {Umanit{\`a}}(2008)}]{fagnola2008}%
  \BibitemOpen
  \bibfield  {author} {\bibinfo {author} {\bibfnamefont {F.}~\bibnamefont
  {Fagnola}}\ and\ \bibinfo {author} {\bibfnamefont {V.}~\bibnamefont
  {Umanit{\`a}}},\ }\href {\doibase 10.1134/S0001434608070092} {\bibfield
  {journal} {\bibinfo  {journal} {Math. Notes}\ }\textbf {\bibinfo {volume}
  {84}},\ \bibinfo {pages} {108} (\bibinfo {year} {2008})}\BibitemShut
  {NoStop}%
\bibitem [{\citenamefont {Jak{\v s}i{\'c}}\ \emph {et~al.}(2014)\citenamefont
  {Jak{\v s}i{\'c}}, \citenamefont {Pillet},\ and\ \citenamefont
  {Westrich}}]{jaksic2014}%
  \BibitemOpen
  \bibfield  {author} {\bibinfo {author} {\bibfnamefont {V.}~\bibnamefont
  {Jak{\v s}i{\'c}}}, \bibinfo {author} {\bibfnamefont {C.~A.}\ \bibnamefont
  {Pillet}}, \ and\ \bibinfo {author} {\bibfnamefont {M.}~\bibnamefont
  {Westrich}},\ }\href {\doibase 10.1007/s10955-013-0826-5} {\bibfield
  {journal} {\bibinfo  {journal} {Journal of Statistical Physics}\ }\textbf
  {\bibinfo {volume} {154}},\ \bibinfo {pages} {153} (\bibinfo {year}
  {2014})}\BibitemShut {NoStop}%
\bibitem [{\citenamefont {Crooks}(2008)}]{crooks2008}%
  \BibitemOpen
  \bibfield  {author} {\bibinfo {author} {\bibfnamefont {G.~E.}\ \bibnamefont
  {Crooks}},\ }\href {\doibase 10.1103/PhysRevA.77.034101} {\bibfield
  {journal} {\bibinfo  {journal} {Phys. Rev. A}\ }\textbf {\bibinfo {volume}
  {77}},\ \bibinfo {pages} {034101} (\bibinfo {year} {2008})}\BibitemShut
  {NoStop}%
\bibitem [{\citenamefont {Roberts}\ \emph {et~al.}(2020)\citenamefont
  {Roberts}, \citenamefont {Lingenfelter},\ and\ \citenamefont
  {Clerk}}]{roberts2020}%
  \BibitemOpen
  \bibfield  {author} {\bibinfo {author} {\bibfnamefont {D.}~\bibnamefont
  {Roberts}}, \bibinfo {author} {\bibfnamefont {A.}~\bibnamefont
  {Lingenfelter}}, \ and\ \bibinfo {author} {\bibfnamefont {A.}~\bibnamefont
  {Clerk}},\ }\href@noop {} {\bibfield  {journal} {\bibinfo  {journal} {arXiv
  preprint arXiv:2011.02148}\ } (\bibinfo {year} {2020})}\BibitemShut {NoStop}%
\bibitem [{\citenamefont {Sieberer}\ \emph {et~al.}(2015)\citenamefont
  {Sieberer}, \citenamefont {Chiocchetta}, \citenamefont {Gambassi},
  \citenamefont {T\"auber},\ and\ \citenamefont {Diehl}}]{sieberer2015}%
  \BibitemOpen
  \bibfield  {author} {\bibinfo {author} {\bibfnamefont {L.~M.}\ \bibnamefont
  {Sieberer}}, \bibinfo {author} {\bibfnamefont {A.}~\bibnamefont
  {Chiocchetta}}, \bibinfo {author} {\bibfnamefont {A.}~\bibnamefont
  {Gambassi}}, \bibinfo {author} {\bibfnamefont {U.~C.}\ \bibnamefont
  {T\"auber}}, \ and\ \bibinfo {author} {\bibfnamefont {S.}~\bibnamefont
  {Diehl}},\ }\href {\doibase 10.1103/PhysRevB.92.134307} {\bibfield  {journal}
  {\bibinfo  {journal} {Phys. Rev. B}\ }\textbf {\bibinfo {volume} {92}},\
  \bibinfo {pages} {134307} (\bibinfo {year} {2015})}\BibitemShut {NoStop}%
\bibitem [{\citenamefont {Benoit}\ \emph {et~al.}(1986)\citenamefont {Benoit},
  \citenamefont {Washburn}, \citenamefont {Umbach}, \citenamefont {Laibowitz},\
  and\ \citenamefont {Webb}}]{benoit1986}%
  \BibitemOpen
  \bibfield  {author} {\bibinfo {author} {\bibfnamefont {A.~D.}\ \bibnamefont
  {Benoit}}, \bibinfo {author} {\bibfnamefont {S.}~\bibnamefont {Washburn}},
  \bibinfo {author} {\bibfnamefont {C.~P.}\ \bibnamefont {Umbach}}, \bibinfo
  {author} {\bibfnamefont {R.~B.}\ \bibnamefont {Laibowitz}}, \ and\ \bibinfo
  {author} {\bibfnamefont {R.~A.}\ \bibnamefont {Webb}},\ }\href {\doibase
  10.1103/PhysRevLett.57.1765} {\bibfield  {journal} {\bibinfo  {journal}
  {Phys. Rev. Lett.}\ }\textbf {\bibinfo {volume} {57}},\ \bibinfo {pages}
  {1765} (\bibinfo {year} {1986})}\BibitemShut {NoStop}%
\bibitem [{\citenamefont {Jacquod}\ \emph {et~al.}(2012)\citenamefont
  {Jacquod}, \citenamefont {Whitney}, \citenamefont {Meair},\ and\
  \citenamefont {B\"uttiker}}]{jacquod2012}%
  \BibitemOpen
  \bibfield  {author} {\bibinfo {author} {\bibfnamefont {P.}~\bibnamefont
  {Jacquod}}, \bibinfo {author} {\bibfnamefont {R.~S.}\ \bibnamefont
  {Whitney}}, \bibinfo {author} {\bibfnamefont {J.}~\bibnamefont {Meair}}, \
  and\ \bibinfo {author} {\bibfnamefont {M.}~\bibnamefont {B\"uttiker}},\
  }\href {\doibase 10.1103/PhysRevB.86.155118} {\bibfield  {journal} {\bibinfo
  {journal} {Phys. Rev. B}\ }\textbf {\bibinfo {volume} {86}},\ \bibinfo
  {pages} {155118} (\bibinfo {year} {2012})}\BibitemShut {NoStop}%
\bibitem [{\citenamefont {Matthews}\ \emph {et~al.}(2014)\citenamefont
  {Matthews}, \citenamefont {Battista}, \citenamefont {S\'anchez},
  \citenamefont {Samuelsson},\ and\ \citenamefont {Linke}}]{matthews2014}%
  \BibitemOpen
  \bibfield  {author} {\bibinfo {author} {\bibfnamefont {J.}~\bibnamefont
  {Matthews}}, \bibinfo {author} {\bibfnamefont {F.}~\bibnamefont {Battista}},
  \bibinfo {author} {\bibfnamefont {D.}~\bibnamefont {S\'anchez}}, \bibinfo
  {author} {\bibfnamefont {P.}~\bibnamefont {Samuelsson}}, \ and\ \bibinfo
  {author} {\bibfnamefont {H.}~\bibnamefont {Linke}},\ }\href {\doibase
  10.1103/PhysRevB.90.165428} {\bibfield  {journal} {\bibinfo  {journal} {Phys.
  Rev. B}\ }\textbf {\bibinfo {volume} {90}},\ \bibinfo {pages} {165428}
  (\bibinfo {year} {2014})}\BibitemShut {NoStop}%
\bibitem [{\citenamefont {L\'opez}\ \emph {et~al.}(2012)\citenamefont
  {L\'opez}, \citenamefont {Lim},\ and\ \citenamefont {S\'anchez}}]{lopez2012}%
  \BibitemOpen
  \bibfield  {author} {\bibinfo {author} {\bibfnamefont {R.}~\bibnamefont
  {L\'opez}}, \bibinfo {author} {\bibfnamefont {J.~S.}\ \bibnamefont {Lim}}, \
  and\ \bibinfo {author} {\bibfnamefont {D.}~\bibnamefont {S\'anchez}},\ }\href
  {\doibase 10.1103/PhysRevLett.108.246603} {\bibfield  {journal} {\bibinfo
  {journal} {Phys. Rev. Lett.}\ }\textbf {\bibinfo {volume} {108}},\ \bibinfo
  {pages} {246603} (\bibinfo {year} {2012})}\BibitemShut {NoStop}%
\bibitem [{\citenamefont {Bruus}\ and\ \citenamefont
  {Flensberg}(2004)}]{bruus2004}%
  \BibitemOpen
  \bibfield  {author} {\bibinfo {author} {\bibfnamefont {H.}~\bibnamefont
  {Bruus}}\ and\ \bibinfo {author} {\bibfnamefont {K.}~\bibnamefont
  {Flensberg}},\ }\href@noop {} {\emph {\bibinfo {title} {Many-body quantum
  theory in condensed matter physics: an introduction}}}\ (\bibinfo
  {publisher} {Oxford university press},\ \bibinfo {year} {2004})\BibitemShut
  {NoStop}%
\bibitem [{\citenamefont {Oka}\ \emph {et~al.}(2014)\citenamefont {Oka},
  \citenamefont {Brovko}, \citenamefont {Corbetta}, \citenamefont {Stepanyuk},
  \citenamefont {Sander},\ and\ \citenamefont {Kirschner}}]{oka2014spin}%
  \BibitemOpen
  \bibfield  {author} {\bibinfo {author} {\bibfnamefont {H.}~\bibnamefont
  {Oka}}, \bibinfo {author} {\bibfnamefont {O.~O.}\ \bibnamefont {Brovko}},
  \bibinfo {author} {\bibfnamefont {M.}~\bibnamefont {Corbetta}}, \bibinfo
  {author} {\bibfnamefont {V.~S.}\ \bibnamefont {Stepanyuk}}, \bibinfo {author}
  {\bibfnamefont {D.}~\bibnamefont {Sander}}, \ and\ \bibinfo {author}
  {\bibfnamefont {J.}~\bibnamefont {Kirschner}},\ }\href {\doibase
  10.1103/RevModPhys.86.1127} {\bibfield  {journal} {\bibinfo  {journal} {Rev.
  Mod. Phys.}\ }\textbf {\bibinfo {volume} {86}},\ \bibinfo {pages} {1127}
  (\bibinfo {year} {2014})}\BibitemShut {NoStop}%
\bibitem [{\citenamefont {Dao}\ \emph {et~al.}(2007)\citenamefont {Dao},
  \citenamefont {Georges}, \citenamefont {Dalibard}, \citenamefont {Salomon},\
  and\ \citenamefont {Carusotto}}]{Dao2007}%
  \BibitemOpen
  \bibfield  {author} {\bibinfo {author} {\bibfnamefont {T.-L.}\ \bibnamefont
  {Dao}}, \bibinfo {author} {\bibfnamefont {A.}~\bibnamefont {Georges}},
  \bibinfo {author} {\bibfnamefont {J.}~\bibnamefont {Dalibard}}, \bibinfo
  {author} {\bibfnamefont {C.}~\bibnamefont {Salomon}}, \ and\ \bibinfo
  {author} {\bibfnamefont {I.}~\bibnamefont {Carusotto}},\ }\href {\doibase
  10.1103/PhysRevLett.98.240402} {\bibfield  {journal} {\bibinfo  {journal}
  {Phys. Rev. Lett.}\ }\textbf {\bibinfo {volume} {98}},\ \bibinfo {pages}
  {240402} (\bibinfo {year} {2007})}\BibitemShut {NoStop}%
\bibitem [{\citenamefont {Pollmann}\ \emph {et~al.}(2010)\citenamefont
  {Pollmann}, \citenamefont {Turner}, \citenamefont {Berg},\ and\ \citenamefont
  {Oshikawa}}]{pollmann2010}%
  \BibitemOpen
  \bibfield  {author} {\bibinfo {author} {\bibfnamefont {F.}~\bibnamefont
  {Pollmann}}, \bibinfo {author} {\bibfnamefont {A.~M.}\ \bibnamefont
  {Turner}}, \bibinfo {author} {\bibfnamefont {E.}~\bibnamefont {Berg}}, \ and\
  \bibinfo {author} {\bibfnamefont {M.}~\bibnamefont {Oshikawa}},\ }\href
  {\doibase 10.1103/PhysRevB.81.064439} {\bibfield  {journal} {\bibinfo
  {journal} {Phys. Rev. B}\ }\textbf {\bibinfo {volume} {81}},\ \bibinfo
  {pages} {064439} (\bibinfo {year} {2010})}\BibitemShut {NoStop}%
\bibitem [{\citenamefont {McGinley}\ and\ \citenamefont
  {Cooper}(2020)}]{mcginley2020}%
  \BibitemOpen
  \bibfield  {author} {\bibinfo {author} {\bibfnamefont {M.}~\bibnamefont
  {McGinley}}\ and\ \bibinfo {author} {\bibfnamefont {N.~R.}\ \bibnamefont
  {Cooper}},\ }\href {https://doi.org/10.1038/s41567-020-0956-z} {\bibfield
  {journal} {\bibinfo  {journal} {Nature Physics}\ } (\bibinfo {year}
  {2020})}\BibitemShut {NoStop}%
\bibitem [{\citenamefont {Diehl}\ \emph {et~al.}(2011)\citenamefont {Diehl},
  \citenamefont {Rico}, \citenamefont {Baranov},\ and\ \citenamefont
  {Zoller}}]{diehl2011}%
  \BibitemOpen
  \bibfield  {author} {\bibinfo {author} {\bibfnamefont {S.}~\bibnamefont
  {Diehl}}, \bibinfo {author} {\bibfnamefont {E.}~\bibnamefont {Rico}},
  \bibinfo {author} {\bibfnamefont {M.~A.}\ \bibnamefont {Baranov}}, \ and\
  \bibinfo {author} {\bibfnamefont {P.}~\bibnamefont {Zoller}},\ }\href
  {\doibase 10.1038/nphys2106} {\bibfield  {journal} {\bibinfo  {journal}
  {Nature Physics}\ }\textbf {\bibinfo {volume} {7}},\ \bibinfo {pages} {971}
  (\bibinfo {year} {2011})}\BibitemShut {NoStop}%
\bibitem [{\citenamefont {Lieu}\ \emph {et~al.}(2020)\citenamefont {Lieu},
  \citenamefont {McGinley},\ and\ \citenamefont {Cooper}}]{lieu2020}%
  \BibitemOpen
  \bibfield  {author} {\bibinfo {author} {\bibfnamefont {S.}~\bibnamefont
  {Lieu}}, \bibinfo {author} {\bibfnamefont {M.}~\bibnamefont {McGinley}}, \
  and\ \bibinfo {author} {\bibfnamefont {N.~R.}\ \bibnamefont {Cooper}},\
  }\href {\doibase 10.1103/PhysRevLett.124.040401} {\bibfield  {journal}
  {\bibinfo  {journal} {Phys. Rev. Lett.}\ }\textbf {\bibinfo {volume} {124}},\
  \bibinfo {pages} {040401} (\bibinfo {year} {2020})}\BibitemShut {NoStop}%
\bibitem [{\citenamefont {van Caspel}\ \emph {et~al.}(2019)\citenamefont {van
  Caspel}, \citenamefont {Arze},\ and\ \citenamefont {Castillo}}]{moos2019}%
  \BibitemOpen
  \bibfield  {author} {\bibinfo {author} {\bibfnamefont {M.}~\bibnamefont {van
  Caspel}}, \bibinfo {author} {\bibfnamefont {S.~E.~T.}\ \bibnamefont {Arze}},
  \ and\ \bibinfo {author} {\bibfnamefont {I.~P.}\ \bibnamefont {Castillo}},\
  }\href@noop {} {\bibfield  {journal} {\bibinfo  {journal} {SciPost Phys}\
  }\textbf {\bibinfo {volume} {6}},\ \bibinfo {pages} {26} (\bibinfo {year}
  {2019})}\BibitemShut {NoStop}%
\bibitem [{\citenamefont {Tonielli}\ \emph {et~al.}(2020)\citenamefont
  {Tonielli}, \citenamefont {Budich}, \citenamefont {Altland},\ and\
  \citenamefont {Diehl}}]{diehl2020}%
  \BibitemOpen
  \bibfield  {author} {\bibinfo {author} {\bibfnamefont {F.}~\bibnamefont
  {Tonielli}}, \bibinfo {author} {\bibfnamefont {J.~C.}\ \bibnamefont
  {Budich}}, \bibinfo {author} {\bibfnamefont {A.}~\bibnamefont {Altland}}, \
  and\ \bibinfo {author} {\bibfnamefont {S.}~\bibnamefont {Diehl}},\ }\href
  {\doibase 10.1103/PhysRevLett.124.240404} {\bibfield  {journal} {\bibinfo
  {journal} {Phys. Rev. Lett.}\ }\textbf {\bibinfo {volume} {124}},\ \bibinfo
  {pages} {240404} (\bibinfo {year} {2020})}\BibitemShut {NoStop}%
\bibitem [{\citenamefont {Gneiting}\ \emph {et~al.}(2020)\citenamefont
  {Gneiting}, \citenamefont {Koottandavida}, \citenamefont {Rozhkov},\ and\
  \citenamefont {Nori}}]{nori2020}%
  \BibitemOpen
  \bibfield  {author} {\bibinfo {author} {\bibfnamefont {C.}~\bibnamefont
  {Gneiting}}, \bibinfo {author} {\bibfnamefont {A.}~\bibnamefont
  {Koottandavida}}, \bibinfo {author} {\bibfnamefont {A.~V.}\ \bibnamefont
  {Rozhkov}}, \ and\ \bibinfo {author} {\bibfnamefont {F.}~\bibnamefont
  {Nori}},\ }\href {https://arxiv.org/abs/2007.05960} {\bibfield  {journal}
  {\bibinfo  {journal} {arXiv:2007.05960}\ } (\bibinfo {year}
  {2020})}\BibitemShut {NoStop}%
\bibitem [{\citenamefont {Song}\ \emph {et~al.}(2019)\citenamefont {Song},
  \citenamefont {Yao},\ and\ \citenamefont {Wang}}]{wang2019}%
  \BibitemOpen
  \bibfield  {author} {\bibinfo {author} {\bibfnamefont {F.}~\bibnamefont
  {Song}}, \bibinfo {author} {\bibfnamefont {S.}~\bibnamefont {Yao}}, \ and\
  \bibinfo {author} {\bibfnamefont {Z.}~\bibnamefont {Wang}},\ }\href {\doibase
  10.1103/PhysRevLett.123.170401} {\bibfield  {journal} {\bibinfo  {journal}
  {Phys. Rev. Lett.}\ }\textbf {\bibinfo {volume} {123}},\ \bibinfo {pages}
  {170401} (\bibinfo {year} {2019})}\BibitemShut {NoStop}%
\bibitem [{\citenamefont {Yoshida}\ \emph {et~al.}(2020)\citenamefont
  {Yoshida}, \citenamefont {Kudo}, \citenamefont {Katsura},\ and\ \citenamefont
  {Hatsugai}}]{yoshida2020}%
  \BibitemOpen
  \bibfield  {author} {\bibinfo {author} {\bibfnamefont {T.}~\bibnamefont
  {Yoshida}}, \bibinfo {author} {\bibfnamefont {K.}~\bibnamefont {Kudo}},
  \bibinfo {author} {\bibfnamefont {H.}~\bibnamefont {Katsura}}, \ and\
  \bibinfo {author} {\bibfnamefont {Y.}~\bibnamefont {Hatsugai}},\ }\href
  {\doibase 10.1103/PhysRevResearch.2.033428} {\bibfield  {journal} {\bibinfo
  {journal} {Phys. Rev. Research}\ }\textbf {\bibinfo {volume} {2}},\ \bibinfo
  {pages} {033428} (\bibinfo {year} {2020})}\BibitemShut {NoStop}%
\bibitem [{\citenamefont {Liu}\ \emph {et~al.}(2020)\citenamefont {Liu},
  \citenamefont {Zhang}, \citenamefont {Yang},\ and\ \citenamefont
  {Chen}}]{chen2020}%
  \BibitemOpen
  \bibfield  {author} {\bibinfo {author} {\bibfnamefont {C.-H.}\ \bibnamefont
  {Liu}}, \bibinfo {author} {\bibfnamefont {K.}~\bibnamefont {Zhang}}, \bibinfo
  {author} {\bibfnamefont {Z.}~\bibnamefont {Yang}}, \ and\ \bibinfo {author}
  {\bibfnamefont {S.}~\bibnamefont {Chen}},\ }\href {\doibase
  10.1103/PhysRevResearch.2.043167} {\bibfield  {journal} {\bibinfo  {journal}
  {Phys. Rev. Research}\ }\textbf {\bibinfo {volume} {2}},\ \bibinfo {pages}
  {043167} (\bibinfo {year} {2020})}\BibitemShut {NoStop}%
\bibitem [{\citenamefont {Okuma}\ and\ \citenamefont {Sato}(2020)}]{okuma2020}%
  \BibitemOpen
  \bibfield  {author} {\bibinfo {author} {\bibfnamefont {N.}~\bibnamefont
  {Okuma}}\ and\ \bibinfo {author} {\bibfnamefont {M.}~\bibnamefont {Sato}},\
  }\href@noop {} {\bibfield  {journal} {\bibinfo  {journal} {arXiv preprint
  arXiv:2011.08175}\ } (\bibinfo {year} {2020})}\BibitemShut {NoStop}%
\bibitem [{\citenamefont {Longhi}(2020)}]{longhi2020}%
  \BibitemOpen
  \bibfield  {author} {\bibinfo {author} {\bibfnamefont {S.}~\bibnamefont
  {Longhi}},\ }\href {\doibase 10.1103/PhysRevB.102.201103} {\bibfield
  {journal} {\bibinfo  {journal} {Phys. Rev. B}\ }\textbf {\bibinfo {volume}
  {102}},\ \bibinfo {pages} {201103} (\bibinfo {year} {2020})}\BibitemShut
  {NoStop}%
\bibitem [{\citenamefont {Pan}\ \emph {et~al.}(2020)\citenamefont {Pan},
  \citenamefont {Li},\ and\ \citenamefont {Gong}}]{pan2020}%
  \BibitemOpen
  \bibfield  {author} {\bibinfo {author} {\bibfnamefont {J.-S.}\ \bibnamefont
  {Pan}}, \bibinfo {author} {\bibfnamefont {L.}~\bibnamefont {Li}}, \ and\
  \bibinfo {author} {\bibfnamefont {J.}~\bibnamefont {Gong}},\ }\href@noop {}
  {\bibfield  {journal} {\bibinfo  {journal} {arXiv preprint arXiv:2010.14862}\
  } (\bibinfo {year} {2020})}\BibitemShut {NoStop}%
\bibitem [{\citenamefont {Sayyad}\ \emph {et~al.}(2020)\citenamefont {Sayyad},
  \citenamefont {Yu}, \citenamefont {Grushin},\ and\ \citenamefont
  {Sieberer}}]{sieberer2020}%
  \BibitemOpen
  \bibfield  {author} {\bibinfo {author} {\bibfnamefont {S.}~\bibnamefont
  {Sayyad}}, \bibinfo {author} {\bibfnamefont {J.}~\bibnamefont {Yu}}, \bibinfo
  {author} {\bibfnamefont {A.~G.}\ \bibnamefont {Grushin}}, \ and\ \bibinfo
  {author} {\bibfnamefont {L.~M.}\ \bibnamefont {Sieberer}},\ }\href@noop {}
  {\bibfield  {journal} {\bibinfo  {journal} {arXiv preprint arXiv:2011.10601}\
  } (\bibinfo {year} {2020})}\BibitemShut {NoStop}%
\bibitem [{\citenamefont {McDonald}\ \emph {et~al.}(2021)\citenamefont
  {McDonald}, \citenamefont {Hanai},\ and\ \citenamefont
  {Clerk}}]{mcdonald2021}%
  \BibitemOpen
  \bibfield  {author} {\bibinfo {author} {\bibfnamefont {A.}~\bibnamefont
  {McDonald}}, \bibinfo {author} {\bibfnamefont {R.}~\bibnamefont {Hanai}}, \
  and\ \bibinfo {author} {\bibfnamefont {A.~A.}\ \bibnamefont {Clerk}},\
  }\href@noop {} {\bibfield  {journal} {\bibinfo  {journal} {arXiv preprint
  arXiv:2103.01941}\ } (\bibinfo {year} {2021})}\BibitemShut {NoStop}%
\bibitem [{\citenamefont {Coser}\ and\ \citenamefont
  {P{\'{e}}rez-Garc{\'{i}}a}(2019)}]{coser2019}%
  \BibitemOpen
  \bibfield  {author} {\bibinfo {author} {\bibfnamefont {A.}~\bibnamefont
  {Coser}}\ and\ \bibinfo {author} {\bibfnamefont {D.}~\bibnamefont
  {P{\'{e}}rez-Garc{\'{i}}a}},\ }\href {\doibase 10.22331/q-2019-08-12-174}
  {\bibfield  {journal} {\bibinfo  {journal} {{Quantum}}\ }\textbf {\bibinfo
  {volume} {3}},\ \bibinfo {pages} {174} (\bibinfo {year} {2019})}\BibitemShut
  {NoStop}%
\bibitem [{\citenamefont {Deng}\ \emph {et~al.}(2020)\citenamefont {Deng},
  \citenamefont {Pan}, \citenamefont {Chen},\ and\ \citenamefont
  {Zhai}}]{deng2020}%
  \BibitemOpen
  \bibfield  {author} {\bibinfo {author} {\bibfnamefont {T.-S.}\ \bibnamefont
  {Deng}}, \bibinfo {author} {\bibfnamefont {L.}~\bibnamefont {Pan}}, \bibinfo
  {author} {\bibfnamefont {Y.}~\bibnamefont {Chen}}, \ and\ \bibinfo {author}
  {\bibfnamefont {H.}~\bibnamefont {Zhai}},\ }\href@noop {} {\bibfield
  {journal} {\bibinfo  {journal} {arXiv:2009.13043}\ } (\bibinfo {year}
  {2020})}\BibitemShut {NoStop}%
\bibitem [{\citenamefont {Kawabata}\ \emph {et~al.}(2019)\citenamefont
  {Kawabata}, \citenamefont {Shiozaki}, \citenamefont {Ueda},\ and\
  \citenamefont {Sato}}]{kawabata2019}%
  \BibitemOpen
  \bibfield  {author} {\bibinfo {author} {\bibfnamefont {K.}~\bibnamefont
  {Kawabata}}, \bibinfo {author} {\bibfnamefont {K.}~\bibnamefont {Shiozaki}},
  \bibinfo {author} {\bibfnamefont {M.}~\bibnamefont {Ueda}}, \ and\ \bibinfo
  {author} {\bibfnamefont {M.}~\bibnamefont {Sato}},\ }\href {\doibase
  10.1103/PhysRevX.9.041015} {\bibfield  {journal} {\bibinfo  {journal} {Phys.
  Rev. X}\ }\textbf {\bibinfo {volume} {9}},\ \bibinfo {pages} {041015}
  (\bibinfo {year} {2019})}\BibitemShut {NoStop}%
\bibitem [{\citenamefont {Leykam}\ \emph {et~al.}(2017)\citenamefont {Leykam},
  \citenamefont {Bliokh}, \citenamefont {Huang}, \citenamefont {Chong},\ and\
  \citenamefont {Nori}}]{nori2017}%
  \BibitemOpen
  \bibfield  {author} {\bibinfo {author} {\bibfnamefont {D.}~\bibnamefont
  {Leykam}}, \bibinfo {author} {\bibfnamefont {K.~Y.}\ \bibnamefont {Bliokh}},
  \bibinfo {author} {\bibfnamefont {C.}~\bibnamefont {Huang}}, \bibinfo
  {author} {\bibfnamefont {Y.~D.}\ \bibnamefont {Chong}}, \ and\ \bibinfo
  {author} {\bibfnamefont {F.}~\bibnamefont {Nori}},\ }\href {\doibase
  10.1103/PhysRevLett.118.040401} {\bibfield  {journal} {\bibinfo  {journal}
  {Phys. Rev. Lett.}\ }\textbf {\bibinfo {volume} {118}},\ \bibinfo {pages}
  {040401} (\bibinfo {year} {2017})}\BibitemShut {NoStop}%
\bibitem [{\citenamefont {Yokomizo}\ and\ \citenamefont
  {Murakami}(2019)}]{murakami2019}%
  \BibitemOpen
  \bibfield  {author} {\bibinfo {author} {\bibfnamefont {K.}~\bibnamefont
  {Yokomizo}}\ and\ \bibinfo {author} {\bibfnamefont {S.}~\bibnamefont
  {Murakami}},\ }\href {\doibase 10.1103/PhysRevLett.123.066404} {\bibfield
  {journal} {\bibinfo  {journal} {Phys. Rev. Lett.}\ }\textbf {\bibinfo
  {volume} {123}},\ \bibinfo {pages} {066404} (\bibinfo {year}
  {2019})}\BibitemShut {NoStop}%
\bibitem [{\citenamefont {Borgnia}\ \emph {et~al.}(2020)\citenamefont
  {Borgnia}, \citenamefont {Kruchkov},\ and\ \citenamefont
  {Slager}}]{borgina2020}%
  \BibitemOpen
  \bibfield  {author} {\bibinfo {author} {\bibfnamefont {D.~S.}\ \bibnamefont
  {Borgnia}}, \bibinfo {author} {\bibfnamefont {A.~J.}\ \bibnamefont
  {Kruchkov}}, \ and\ \bibinfo {author} {\bibfnamefont {R.-J.}\ \bibnamefont
  {Slager}},\ }\href {\doibase 10.1103/PhysRevLett.124.056802} {\bibfield
  {journal} {\bibinfo  {journal} {Phys. Rev. Lett.}\ }\textbf {\bibinfo
  {volume} {124}},\ \bibinfo {pages} {056802} (\bibinfo {year}
  {2020})}\BibitemShut {NoStop}%
\bibitem [{\citenamefont {Yao}\ and\ \citenamefont {Wang}(2018)}]{yao2018}%
  \BibitemOpen
  \bibfield  {author} {\bibinfo {author} {\bibfnamefont {S.}~\bibnamefont
  {Yao}}\ and\ \bibinfo {author} {\bibfnamefont {Z.}~\bibnamefont {Wang}},\
  }\href {\doibase 10.1103/PhysRevLett.121.086803} {\bibfield  {journal}
  {\bibinfo  {journal} {Phys. Rev. Lett.}\ }\textbf {\bibinfo {volume} {121}},\
  \bibinfo {pages} {086803} (\bibinfo {year} {2018})}\BibitemShut {NoStop}%
\bibitem [{\citenamefont {Malzard}\ \emph {et~al.}(2015)\citenamefont
  {Malzard}, \citenamefont {Poli},\ and\ \citenamefont
  {Schomerus}}]{malzard2015}%
  \BibitemOpen
  \bibfield  {author} {\bibinfo {author} {\bibfnamefont {S.}~\bibnamefont
  {Malzard}}, \bibinfo {author} {\bibfnamefont {C.}~\bibnamefont {Poli}}, \
  and\ \bibinfo {author} {\bibfnamefont {H.}~\bibnamefont {Schomerus}},\ }\href
  {\doibase 10.1103/PhysRevLett.115.200402} {\bibfield  {journal} {\bibinfo
  {journal} {Phys. Rev. Lett.}\ }\textbf {\bibinfo {volume} {115}},\ \bibinfo
  {pages} {200402} (\bibinfo {year} {2015})}\BibitemShut {NoStop}%
\bibitem [{\citenamefont {Herviou}\ \emph {et~al.}(2019)\citenamefont
  {Herviou}, \citenamefont {Regnault},\ and\ \citenamefont
  {Bardarson}}]{loic2019}%
  \BibitemOpen
  \bibfield  {author} {\bibinfo {author} {\bibfnamefont {L.}~\bibnamefont
  {Herviou}}, \bibinfo {author} {\bibfnamefont {N.}~\bibnamefont {Regnault}}, \
  and\ \bibinfo {author} {\bibfnamefont {J.~H.}\ \bibnamefont {Bardarson}},\
  }\href {\doibase 10.21468/SciPostPhys.7.5.069} {\bibfield  {journal}
  {\bibinfo  {journal} {SciPost Phys.}\ }\textbf {\bibinfo {volume} {7}},\
  \bibinfo {pages} {69} (\bibinfo {year} {2019})}\BibitemShut {NoStop}%
\bibitem [{\citenamefont {Lieu}(2018)}]{lieu2018}%
  \BibitemOpen
  \bibfield  {author} {\bibinfo {author} {\bibfnamefont {S.}~\bibnamefont
  {Lieu}},\ }\href {\doibase 10.1103/PhysRevB.98.115135} {\bibfield  {journal}
  {\bibinfo  {journal} {Phys. Rev. B}\ }\textbf {\bibinfo {volume} {98}},\
  \bibinfo {pages} {115135} (\bibinfo {year} {2018})}\BibitemShut {NoStop}%
\bibitem [{\citenamefont {Sato}\ \emph {et~al.}(2012)\citenamefont {Sato},
  \citenamefont {Hasebe}, \citenamefont {Esaki},\ and\ \citenamefont
  {Kohmoto}}]{sato2012}%
  \BibitemOpen
  \bibfield  {author} {\bibinfo {author} {\bibfnamefont {M.}~\bibnamefont
  {Sato}}, \bibinfo {author} {\bibfnamefont {K.}~\bibnamefont {Hasebe}},
  \bibinfo {author} {\bibfnamefont {K.}~\bibnamefont {Esaki}}, \ and\ \bibinfo
  {author} {\bibfnamefont {M.}~\bibnamefont {Kohmoto}},\ }\href@noop {}
  {\bibfield  {journal} {\bibinfo  {journal} {Prog. Theor. Phys.}\ }\textbf
  {\bibinfo {volume} {127}},\ \bibinfo {pages} {937} (\bibinfo {year}
  {2012})}\BibitemShut {NoStop}%
\bibitem [{\citenamefont {Zhou}\ and\ \citenamefont {Lee}(2019)}]{zhou2019}%
  \BibitemOpen
  \bibfield  {author} {\bibinfo {author} {\bibfnamefont {H.}~\bibnamefont
  {Zhou}}\ and\ \bibinfo {author} {\bibfnamefont {J.~Y.}\ \bibnamefont {Lee}},\
  }\href {\doibase 10.1103/PhysRevB.99.235112} {\bibfield  {journal} {\bibinfo
  {journal} {Phys. Rev. B}\ }\textbf {\bibinfo {volume} {99}},\ \bibinfo
  {pages} {235112} (\bibinfo {year} {2019})}\BibitemShut {NoStop}%
\bibitem [{\citenamefont {Kondo}\ \emph {et~al.}(2020)\citenamefont {Kondo},
  \citenamefont {Akagi},\ and\ \citenamefont {Katsura}}]{kondo2020}%
  \BibitemOpen
  \bibfield  {author} {\bibinfo {author} {\bibfnamefont {H.}~\bibnamefont
  {Kondo}}, \bibinfo {author} {\bibfnamefont {Y.}~\bibnamefont {Akagi}}, \ and\
  \bibinfo {author} {\bibfnamefont {H.}~\bibnamefont {Katsura}},\ }\href@noop
  {} {\bibfield  {journal} {\bibinfo  {journal} {arXiv preprint
  arXiv:2006.10391}\ } (\bibinfo {year} {2020})}\BibitemShut {NoStop}%
\bibitem [{\citenamefont {Bergholtz}\ \emph {et~al.}(2019)\citenamefont
  {Bergholtz}, \citenamefont {Budich},\ and\ \citenamefont
  {Kunst}}]{bergholtz2019}%
  \BibitemOpen
  \bibfield  {author} {\bibinfo {author} {\bibfnamefont {E.~J.}\ \bibnamefont
  {Bergholtz}}, \bibinfo {author} {\bibfnamefont {J.~C.}\ \bibnamefont
  {Budich}}, \ and\ \bibinfo {author} {\bibfnamefont {F.~K.}\ \bibnamefont
  {Kunst}},\ }\href {https://arxiv.org/abs/1912.10048} {\bibfield  {journal}
  {\bibinfo  {journal} {arXiv:1912.10048}\ } (\bibinfo {year}
  {2019})}\BibitemShut {NoStop}%
\bibitem [{\citenamefont {Lindblad}(1976)}]{lindblad1976}%
  \BibitemOpen
  \bibfield  {author} {\bibinfo {author} {\bibfnamefont {G.}~\bibnamefont
  {Lindblad}},\ }\href {https://link.springer.com/article/10.1007/BF01608499}
  {\bibfield  {journal} {\bibinfo  {journal} {Commun. Math. Phys}\ }\textbf
  {\bibinfo {volume} {48}},\ \bibinfo {pages} {119} (\bibinfo {year}
  {1976})}\BibitemShut {NoStop}%
\bibitem [{\citenamefont {Gorini}\ \emph {et~al.}(1976)\citenamefont {Gorini},
  \citenamefont {Kossakowski},\ and\ \citenamefont {Sudarshan}}]{gorini1976}%
  \BibitemOpen
  \bibfield  {author} {\bibinfo {author} {\bibfnamefont {V.}~\bibnamefont
  {Gorini}}, \bibinfo {author} {\bibfnamefont {A.}~\bibnamefont {Kossakowski}},
  \ and\ \bibinfo {author} {\bibfnamefont {E.~C.~G.}\ \bibnamefont
  {Sudarshan}},\ }\href {\doibase 10.1063/1.522979} {\bibfield  {journal}
  {\bibinfo  {journal} {J. Math. Phys.}\ }\textbf {\bibinfo {volume} {17}},\
  \bibinfo {pages} {821} (\bibinfo {year} {1976})}\BibitemShut {NoStop}%
\bibitem [{\citenamefont {Breuer}\ and\ \citenamefont
  {Petruccione}(2002)}]{breuer2002}%
  \BibitemOpen
  \bibfield  {author} {\bibinfo {author} {\bibfnamefont {H.}~\bibnamefont
  {Breuer}}\ and\ \bibinfo {author} {\bibfnamefont {F.}~\bibnamefont
  {Petruccione}},\ }\href {https://books.google.co.uk/books?id=w2UOnwEACAAJ}
  {\emph {\bibinfo {title} {The Theory of Open Quantum Systems}}}\ (\bibinfo
  {publisher} {Oxford University Press},\ \bibinfo {year} {2002})\BibitemShut
  {NoStop}%
\bibitem [{\citenamefont {Altland}\ \emph {et~al.}(2020)\citenamefont
  {Altland}, \citenamefont {Fleischhauer},\ and\ \citenamefont
  {Diehl}}]{altland2020}%
  \BibitemOpen
  \bibfield  {author} {\bibinfo {author} {\bibfnamefont {A.}~\bibnamefont
  {Altland}}, \bibinfo {author} {\bibfnamefont {M.}~\bibnamefont
  {Fleischhauer}}, \ and\ \bibinfo {author} {\bibfnamefont {S.}~\bibnamefont
  {Diehl}},\ }\href {https://arxiv.org/abs/2007.10448} {\bibfield  {journal}
  {\bibinfo  {journal} {arXiv:2007.10448}\ } (\bibinfo {year}
  {2020})}\BibitemShut {NoStop}%
\bibitem [{Note1()}]{Note1}%
  \BibitemOpen
  \bibinfo {note} {The phase $\theta $ carry no physical significance, since
  jump operators can be changed by a transform $L_i \rightarrow e^{ i \chi
  }L_i$ without changing the Lindbladian \protect \textup {\hbox {\mathsurround
  \z@ \protect \normalfont (\ignorespaces \ref {eq:lindbladeq}\unskip
  \@@italiccorr )}}.}\BibitemShut {Stop}%
\bibitem [{SM()}]{SM}%
  \BibitemOpen
  \href@noop {} {}\bibinfo {note} {See the Supplemental Material for ***.
  Contains Refs.~***}\BibitemShut {NoStop}%
\bibitem [{\citenamefont {Ryu}\ \emph {et~al.}(2010)\citenamefont {Ryu},
  \citenamefont {Schnyder}, \citenamefont {Furusaki},\ and\ \citenamefont
  {Ludwig}}]{ryu2010}%
  \BibitemOpen
  \bibfield  {author} {\bibinfo {author} {\bibfnamefont {S.}~\bibnamefont
  {Ryu}}, \bibinfo {author} {\bibfnamefont {A.~P.}\ \bibnamefont {Schnyder}},
  \bibinfo {author} {\bibfnamefont {A.}~\bibnamefont {Furusaki}}, \ and\
  \bibinfo {author} {\bibfnamefont {A.~W.~W.}\ \bibnamefont {Ludwig}},\ }\href
  {\doibase 10.1088/1367-2630/12/6/065010} {\bibfield  {journal} {\bibinfo
  {journal} {New J. Phys.}\ }\textbf {\bibinfo {volume} {12}},\ \bibinfo
  {pages} {065010} (\bibinfo {year} {2010})}\BibitemShut {NoStop}%
\bibitem [{\citenamefont {Alicki}\ \emph {et~al.}(2009)\citenamefont {Alicki},
  \citenamefont {Fannes},\ and\ \citenamefont {Horodecki}}]{alicki2009}%
  \BibitemOpen
  \bibfield  {author} {\bibinfo {author} {\bibfnamefont {R.}~\bibnamefont
  {Alicki}}, \bibinfo {author} {\bibfnamefont {M.}~\bibnamefont {Fannes}}, \
  and\ \bibinfo {author} {\bibfnamefont {M.}~\bibnamefont {Horodecki}},\ }\href
  {\doibase 10.1088/1751-8113/42/6/065303} {\bibfield  {journal} {\bibinfo
  {journal} {Journal of Physics A: Mathematical and Theoretical}\ }\textbf
  {\bibinfo {volume} {42}},\ \bibinfo {pages} {065303} (\bibinfo {year}
  {2009})}\BibitemShut {NoStop}%
\bibitem [{\citenamefont {Bravyi}\ and\ \citenamefont
  {Haah}(2013)}]{bravyi2013}%
  \BibitemOpen
  \bibfield  {author} {\bibinfo {author} {\bibfnamefont {S.}~\bibnamefont
  {Bravyi}}\ and\ \bibinfo {author} {\bibfnamefont {J.}~\bibnamefont {Haah}},\
  }\href {\doibase 10.1103/PhysRevLett.111.200501} {\bibfield  {journal}
  {\bibinfo  {journal} {Phys. Rev. Lett.}\ }\textbf {\bibinfo {volume} {111}},\
  \bibinfo {pages} {200501} (\bibinfo {year} {2013})}\BibitemShut {NoStop}%
\bibitem [{\citenamefont {Kunst}\ \emph {et~al.}(2018)\citenamefont {Kunst},
  \citenamefont {Edvardsson}, \citenamefont {Budich},\ and\ \citenamefont
  {Bergholtz}}]{kunst2018}%
  \BibitemOpen
  \bibfield  {author} {\bibinfo {author} {\bibfnamefont {F.~K.}\ \bibnamefont
  {Kunst}}, \bibinfo {author} {\bibfnamefont {E.}~\bibnamefont {Edvardsson}},
  \bibinfo {author} {\bibfnamefont {J.~C.}\ \bibnamefont {Budich}}, \ and\
  \bibinfo {author} {\bibfnamefont {E.~J.}\ \bibnamefont {Bergholtz}},\ }\href
  {\doibase 10.1103/PhysRevLett.121.026808} {\bibfield  {journal} {\bibinfo
  {journal} {Phys. Rev. Lett.}\ }\textbf {\bibinfo {volume} {121}},\ \bibinfo
  {pages} {026808} (\bibinfo {year} {2018})}\BibitemShut {NoStop}%
\bibitem [{\citenamefont {Denisov}\ \emph {et~al.}(2019)\citenamefont
  {Denisov}, \citenamefont {Laptyeva}, \citenamefont {Tarnowski}, \citenamefont
  {Chru\ifmmode \acute{s}\else \'{s}\fi{}ci\ifmmode~\acute{n}\else
  \'{n}\fi{}ski},\ and\ \citenamefont {\ifmmode~\dot{Z}\else
  \.{Z}\fi{}yczkowski}}]{denisov2019}%
  \BibitemOpen
  \bibfield  {author} {\bibinfo {author} {\bibfnamefont {S.}~\bibnamefont
  {Denisov}}, \bibinfo {author} {\bibfnamefont {T.}~\bibnamefont {Laptyeva}},
  \bibinfo {author} {\bibfnamefont {W.}~\bibnamefont {Tarnowski}}, \bibinfo
  {author} {\bibfnamefont {D.}~\bibnamefont {Chru\ifmmode \acute{s}\else
  \'{s}\fi{}ci\ifmmode~\acute{n}\else \'{n}\fi{}ski}}, \ and\ \bibinfo {author}
  {\bibfnamefont {K.}~\bibnamefont {\ifmmode~\dot{Z}\else
  \.{Z}\fi{}yczkowski}},\ }\href {\doibase 10.1103/PhysRevLett.123.140403}
  {\bibfield  {journal} {\bibinfo  {journal} {Phys. Rev. Lett.}\ }\textbf
  {\bibinfo {volume} {123}},\ \bibinfo {pages} {140403} (\bibinfo {year}
  {2019})}\BibitemShut {NoStop}%
\bibitem [{\citenamefont {Can}\ \emph {et~al.}(2019)\citenamefont {Can},
  \citenamefont {Oganesyan}, \citenamefont {Orgad},\ and\ \citenamefont
  {Gopalakrishnan}}]{can2019}%
  \BibitemOpen
  \bibfield  {author} {\bibinfo {author} {\bibfnamefont {T.}~\bibnamefont
  {Can}}, \bibinfo {author} {\bibfnamefont {V.}~\bibnamefont {Oganesyan}},
  \bibinfo {author} {\bibfnamefont {D.}~\bibnamefont {Orgad}}, \ and\ \bibinfo
  {author} {\bibfnamefont {S.}~\bibnamefont {Gopalakrishnan}},\ }\href
  {\doibase 10.1103/PhysRevLett.123.234103} {\bibfield  {journal} {\bibinfo
  {journal} {Phys. Rev. Lett.}\ }\textbf {\bibinfo {volume} {123}},\ \bibinfo
  {pages} {234103} (\bibinfo {year} {2019})}\BibitemShut {NoStop}%
\bibitem [{\citenamefont {S\'a}\ \emph {et~al.}(2020)\citenamefont {S\'a},
  \citenamefont {Ribeiro},\ and\ \citenamefont {Prosen}}]{prosen2020}%
  \BibitemOpen
  \bibfield  {author} {\bibinfo {author} {\bibfnamefont {L.}~\bibnamefont
  {S\'a}}, \bibinfo {author} {\bibfnamefont {P.}~\bibnamefont {Ribeiro}}, \
  and\ \bibinfo {author} {\bibfnamefont {T.~c.~v.}\ \bibnamefont {Prosen}},\
  }\href {\doibase 10.1103/PhysRevX.10.021019} {\bibfield  {journal} {\bibinfo
  {journal} {Phys. Rev. X}\ }\textbf {\bibinfo {volume} {10}},\ \bibinfo
  {pages} {021019} (\bibinfo {year} {2020})}\BibitemShut {NoStop}%
\bibitem [{\citenamefont {Wang}\ \emph {et~al.}(2020)\citenamefont {Wang},
  \citenamefont {Piazza},\ and\ \citenamefont {Luitz}}]{wang2020}%
  \BibitemOpen
  \bibfield  {author} {\bibinfo {author} {\bibfnamefont {K.}~\bibnamefont
  {Wang}}, \bibinfo {author} {\bibfnamefont {F.}~\bibnamefont {Piazza}}, \ and\
  \bibinfo {author} {\bibfnamefont {D.~J.}\ \bibnamefont {Luitz}},\ }\href
  {\doibase 10.1103/PhysRevLett.124.100604} {\bibfield  {journal} {\bibinfo
  {journal} {Phys. Rev. Lett.}\ }\textbf {\bibinfo {volume} {124}},\ \bibinfo
  {pages} {100604} (\bibinfo {year} {2020})}\BibitemShut {NoStop}%
\bibitem [{\citenamefont {Hamazaki}\ \emph {et~al.}(2020)\citenamefont
  {Hamazaki}, \citenamefont {Kawabata}, \citenamefont {Kura},\ and\
  \citenamefont {Ueda}}]{hamazaki2020}%
  \BibitemOpen
  \bibfield  {author} {\bibinfo {author} {\bibfnamefont {R.}~\bibnamefont
  {Hamazaki}}, \bibinfo {author} {\bibfnamefont {K.}~\bibnamefont {Kawabata}},
  \bibinfo {author} {\bibfnamefont {N.}~\bibnamefont {Kura}}, \ and\ \bibinfo
  {author} {\bibfnamefont {M.}~\bibnamefont {Ueda}},\ }\href {\doibase
  10.1103/PhysRevResearch.2.023286} {\bibfield  {journal} {\bibinfo  {journal}
  {Phys. Rev. Research}\ }\textbf {\bibinfo {volume} {2}},\ \bibinfo {pages}
  {023286} (\bibinfo {year} {2020})}\BibitemShut {NoStop}%
\bibitem [{\citenamefont {Prosen}(2008)}]{prosen2008}%
  \BibitemOpen
  \bibfield  {author} {\bibinfo {author} {\bibfnamefont {T.}~\bibnamefont
  {Prosen}},\ }\href
  {https://iopscience.iop.org/article/10.1088/1367-2630/10/4/043026/meta}
  {\bibfield  {journal} {\bibinfo  {journal} {New J. Phys.}\ }\textbf {\bibinfo
  {volume} {10}},\ \bibinfo {pages} {043026} (\bibinfo {year}
  {2008})}\BibitemShut {NoStop}%
\bibitem [{\citenamefont {Bu{\v{c}}a}\ and\ \citenamefont
  {Prosen}(2012)}]{buca2012}%
  \BibitemOpen
  \bibfield  {author} {\bibinfo {author} {\bibfnamefont {B.}~\bibnamefont
  {Bu{\v{c}}a}}\ and\ \bibinfo {author} {\bibfnamefont {T.}~\bibnamefont
  {Prosen}},\ }\href {\doibase 10.1088/1367-2630/14/7/073007} {\bibfield
  {journal} {\bibinfo  {journal} {New J. Phys.}\ }\textbf {\bibinfo {volume}
  {14}},\ \bibinfo {pages} {073007} (\bibinfo {year} {2012})}\BibitemShut
  {NoStop}%
\bibitem [{\citenamefont {Kubo}(1957)}]{kubo1957}%
  \BibitemOpen
  \bibfield  {author} {\bibinfo {author} {\bibfnamefont {R.}~\bibnamefont
  {Kubo}},\ }\href@noop {} {\bibfield  {journal} {\bibinfo  {journal} {J. Phys.
  Soc. Jpn}\ }\textbf {\bibinfo {volume} {12}},\ \bibinfo {pages} {570}
  (\bibinfo {year} {1957})}\BibitemShut {NoStop}%
\end{thebibliography}%
\bibliographystyle{apsrev4-1}

\newpage
\afterpage{\blankpage}

\newpage
\widetext

\setcounter{equation}{0}
\setcounter{figure}{0}
\setcounter{table}{0}
\setcounter{page}{1}

\renewcommand{\theequation}{S\arabic{equation}}
\renewcommand{\thefigure}{S\arabic{figure}}

\begin{center}
		{\fontsize{12}{12}\selectfont
			\textbf{Supplemental Material for  ``\papertitle''\\[5mm]}}
		
\end{center}
\normalsize\

The Supplemental Material is organized as follows: In Sec.~1, we complete the steps in the proof of the Kramers’ theorem for Lindbladians. We also explain why the degeneracy only appears for fermonic systems (as opposed to spin systems) and describe an analogous Kramers’ degeneracy for thermal quantum channels. In Sec.~2, we show how the Kramers’ degeneracy manifests in the single-particle spectrum of ``quadratic Lindbladians''. Sec.~3 shows that microreversibility of the Lindbladian arises naturally from TRS-invariant system-bath coupling in the case of a thermal bath. Sec.~4 describes linear response in open quantum systems, suggesting that the degeneracy can be probed via tunneling spectroscopy experiments.

\section{ 1.~Kramers' theorem and Green's function degeneracy}

\textbf{Orthogonal solutions}.~We complete the steps of the generalized Kramers' theorem outlined in the main text. From the main text, we have seen that if the Lindbladian satisfies a microreversibility condition: $\hL^\dagger_- =   \hQ^{-1}_-  \hT_- ^{-1} \hL_-  \hT_-  \hQ_-$ with $\hL_-(r_i) = \Lambda_i r_i, \hL_-^\dagger(l_i) = \Lambda_i^* l_i,$ then $r_i  $ and $ \hT \hQ (l_i)  $ are both right eigenoperators of $ \hL_-$ with eigenvalue $ \Lambda_i$. We now proceed to show that these are indeed orthogonal solutions by showing that $\tr[l_i^\dagger    \hT \hQ  (l_i) ]=0$:
\begin{align}
\tr[l_i^\dagger T(q l_i)T^{-1} ] &= \tr [ ( T(T (q l_i)T^{-1})T^{-1})^\dagger T l_i T^{-1} ] \label{eq:first} \\ 
& = \tr [ ( T^2 (q l_i)T^{-2})^\dagger T l_i T^{-1} ] \\
& = - \tr [ ( q l_i)^\dagger T l_i T^{-1} ] \\ \label{eq:par}
& = - \tr [  l_i^\dagger q  T l_i T^{-1} ] \\
& = - \tr [  l_i^\dagger  T (q l_i) T^{-1} ] \\
& = 0.
\label{eq:orthog}
\end{align}
In Eq.~\eqref{eq:first}, we have used the relation: 
\begin{align}
\tr[ (T \psi T^{-1})^\dagger (T \phi T^{-1})] &= \tr[ (U \psi^* U^\dagger)^\dagger (U \phi^* U^\dagger) ] \\
&= \tr[ U \psi^T U^\dagger U \phi^* U^\dagger ] \\
&= \tr[ \psi^T \phi^*  ] \\
& =  \tr[ \phi^\dagger \psi  ], 
\end{align}
for any $\psi, \phi$, where $T = U K$ . In Eq.~\eqref{eq:par}, we have used $\hT^2( q l_i) = q \hT^2(  l_i) = -q  l_i$, i.e.~$l_i$ is an eigenoperator of the odd-superparity sector of the Lindbladian.

\textbf{Definition of the Green's function}. Here we highlight an important subtlety regarding the definition of the Green's function Eq.~\eqref{eq:retardedgreens}, and in particular the interpretation of the time-evolved fermion operator $\hat{f}_{i, \sigma}(t)$.

A general Green's function of two observables $G_{AB}(t) = \Tr(A(t) B \rho_{\rm SS})$ describes the influence of a perturbation at time $0$ on the outcome of a measurement at time $t$. Since fermion parity symmetry is fundamental, it is not possible to perturb the system by a fermion-parity-odd operator such as $f_{i,\sigma}^\dagger$; rather, in any physical protocol where the Green's function is measured, the initial perturbation will involve some exchange of fermions between the system and some probe. Thus, we should understand that the operator $B$ is a product of $\hat{f}_{i,\sigma}$ with some fermionic operator acting on the probe. (See Section 4 of the SM for an example of this construction in the context of tunneling spectroscopy.) We heuristically write $B \sim f_{i,\sigma}^\dagger f_{\rm pr}$, where $f_{\rm probe}$ is an unspecified fermionic operator acting on the probe, ensuring that $B$ itself is a superparity-even operator. Similarly, the observable $A$ that we measure at time $t$ will be a product of fermion-superparity-odd operators on the system and probe: $A \sim f_{i,\sigma} f_{\rm pr}'$ for some other probe operator $f_{\rm pr}'$.

With this in mind, the Heisenberg picture evolution of the observable  $A$ is given by $A(t) = e^{\mathcal{L}^\dagger t}[A]$, where $\mathcal{L}^\dagger$ is defined in Eq.~\eqref{eq:adjoint}. The Lindbladian only involves operators acting on the system, and not the probe. Therefore, when considering the observable $A = f_{i,\sigma} f_{\rm pr}'$, it is tempting to pull the probe operator $f_{\rm pr}'$ outside the evolution superoperator, so that it can combine with the other probe operator in $B$. Indeed, this is precisely what is done when calculating fermionic Green's functions in closed systems where $\mathcal{L} = -i[H, \cdot]$. However, when the system is open, it is possible for fermions to move between the system and the environment (not to be confused with the probe), which leads to jump operators $L_i$ that are superparity odd. In this case, we cannot ignore the presence of the probe operators, because $\mathcal{L}^\dagger[A]$ will include a term $L_i^\dagger f_{i,\sigma} f_{\rm  pr}' L_i$, which differs from the na{\"i}ve expression $(L_i^\dagger f_{i,\sigma} L_i) f_{\rm  pr}'$ by a minus sign. In other words, even though $f_{\rm pr}'$ is a probe operator, its anticommutation with the jump operators means that the evolution of the product $f_{i,\sigma} f_{\rm  pr}'$ doesn't factorize as the product of time-evolved operators $f_{i,\sigma}(t)$ and $f_{\rm  pr}'$.

This can be remedied by defining a `dummy' Majorana fermion operator $\eta_{\rm d}$, having no dynamics of its own, and including it such that the correct anticommutation relations are obeyed. Specifically, one can verify that $\mathcal{L}^\dagger[f_{i,\sigma} f_{\rm  pr}'] = \eta_{\rm d}\, \mathcal{L}^\dagger[\eta_{\rm d} f_{i,\sigma} ]  f_{\rm pr}' $, since $\eta_{\rm d}^2 = 1$. This allows probe operators to be pulled out of the system evolution superoperator, such that Green's functions can be defined using operators that pertain to the system only. Specifically, the physically meaningful definition of the time-evolved operator $f_{i,\sigma}(t)$ appearing in Eq.~\eqref{eq:retardedgreens} should be:
\begin{align}\label{eqs:dummy_majorana}
    f_{i,\sigma}(t) \coloneqq \eta_{\rm d}\, e^{\mathcal{L}^\dagger t}[\eta_{\rm d} f_{i,\sigma} ]. 
\end{align}
In practice, we will not explicitly write out the dummy Majorana fermion, but instead understand that it is implicitly included in any expression of the form $e^{\mathcal{L}^\dagger t}[f_{i,\sigma}]$. Alternatively, we can modify all the fermion-superparity-odd jump operators by $L_i \rightarrow \eta_{\rm d} L_i $, in which case the usual expression $e^{\mathcal{L}^\dagger t}[f_{i,\sigma}]$ can be used without modification.

\textbf{Degenerate Green's functions}.~Having dealt with the above issue, we now show that steady-state Green's functions  corresponding to fermions with opposite spin  are related in a simple way due to microreversibility. We first consider the quantity 
\begin{align}
\tr[ e^{\hL^\dagger t} (f_{i, \sigma}) f_{j, \tau}^\dagger q ] &= \tr[\hQ( e^{\hL^\dagger t}(f_{i,\sigma}) f_{j,\tau}^\dagger ) ] \\
& = \tr[ \hT \hQ ( e^{\hL^\dagger t}(f_{i,\sigma}) f_{j,\tau}^\dagger ) ]^* \label{eq:s1} \\
& = \tr[ \hT \hQ ( e^{\hL^\dagger t}(f_{i,\sigma}) ) \hT ( f_{j,\tau}^\dagger ) ]^*  \label{eq:s2} \\
& = \tr[  e^{\hL t} ( \hT \hQ (f_{i,\sigma}))  \hT ( f_{j,\tau}^\dagger ) ]^*\label{eq:s3} \\
& = \sigma \tau \tr[  e^{\hL t} ( q f_{i,-\sigma})  f_{j,-\tau}^\dagger  ]^* \label{eq:s4} \\
& = \sigma \tau \tr[   ( q f_{i,-\sigma}) e^{\hL^\dagger t}  (f_{j,-\tau}^\dagger)  ]^* \label{eq:s5},
\end{align}
where we have used $\tr[A] = \tr[\hT(A)]^*$ in \eqref{eq:s1},  $\hT[ A B] = \hT[A] \hT[B]$ in \eqref{eq:s2}, the definition of microreversibility in \eqref{eq:s3}, $\hT[f_{i,\sigma}] = \sigma f_{i,-\sigma}$ in \eqref{eq:s4}, and $\tr[ A e^{\hL t}  (B) ] = \tr[ e^{\hL^\dagger t}(A) B]$ in \eqref{eq:s5}.

Let us now define the following generalizations of retarded Green's functions:
\begin{align}
G_{i \sigma; j \tau} &\equiv -i \Theta(t) \left(  \tr[  e^{\hL^\dagger t} (f_{i, \sigma}) f_{j, \tau}^\dagger q ] +  \tr[  f_{j, \tau}^\dagger e^{\hL^\dagger t} (f_{i, \sigma})  q ] \right) \\
&= -i \Theta(t) \left( \sigma \tau \tr[f_{i,-\sigma} e^{\hL^\dagger t}(f_{j,-\tau}^\dagger) q]^* + \sigma \tau \tr[e^{\hL^\dagger t}(f_{j,-\tau}^\dagger ) f_{i,-\sigma} q ]^*  \right) \\
&=  -i \Theta(t) \left( \sigma \tau \tr[ e^{\hL^\dagger t}(f_{j,-\tau})  f_{i,-\sigma}^\dagger q] + \sigma \tau \tr[e^{\hL^\dagger t}(f_{j,-\tau} )  q  f_{i,-\sigma}^\dagger] \label{eq:s6} \right) \\
&=  -i \Theta(t) \left( \sigma \tau G_{j -\tau; i -\sigma}(t) \right),
\label{eq:GreenDegeneracy}
\end{align}
where in \eqref{eq:s6} we have used: $\tr[ A B ] = \tr[A^\dagger B^\dagger]^*$ and $[e^{\hL^\dagger t} (A )]^\dagger = e^{\hL^\dagger t} (A^\dagger)$. So indeed we find that steady-state Green's functions for opposite spin labels are related to each other in a simple way for systems with microreversibility. We have confirmed these expressions numerically for the example system described in the main text. 

\textbf{Non-fermionic systems}.~As mentioned in the main text, the Kramers' degeneracy of the Lindbladian only arises in fermionic systems, but not in bosonic or spin systems, even though the latter have a Kramers-degenerate Hamiltonian when the total spin is a half-integer. The differences between fermionic vs.~non-fermionic open systems becomes apparent when we consider how linear response functions are constrained by microreversibility and TRS, in analogy to Eq.~\eqref{eq:GreenDegeneracy}. Regardless of particle statistics, detailed balance implies [Ref.~\cite{agarwal1973}, Eq.~(2.2)]:
\begin{align}
    \Tr[ A(t) B q] = \Tr[ \tilde{B}(t) \tilde{A} q], 
    \label{eq:flucdissAB}
\end{align}
where $\tilde{A} \coloneqq \mathcal{T}[A]^\dagger$ and similar for $\tilde{B}$. Suppose for the sake of simplicity that $\beta = 0$ and hence $q \propto \mathbb{I}$. Then, by decomposing $A$ and $B$ in terms of right and left eigenoperators of $\mathcal{L}$, respectively, the left-hand side of the above can be written as a linear combination of terms $\Tr[r_i(t) l_j] = e^{\Lambda_i t} \delta_{ij}$, and thus without loss of generality the condition becomes $\Tr[r_i(t) l_i] = \Tr[\tilde{l}_i(t) \tilde{r}_i]$. (The same arguments can in principle be generalized to $\beta > 0$ by defining a new set of operators $l_i'$ which are left eigenoperators of $\mathcal{L}$ with respect to the non-standard inner product $\langle A, B \rangle_q = \Tr[A^\dagger B q]$.)

In a spin system, we find that $\tilde{l}_i = r_i$, which ensures that the above condition can be satisfied, regardless of whether or not $\mathcal{L}$ possesses any degeneracies. The same relation cannot hold true for fermionic superparity-odd eigenoperators, because it would contradict with Eq.~\eqref{eq:orthog} (when one sets $q \propto \mathbb{I}$). The only way for the Green's function identity to hold is if $r_i$ and $\tilde{l}_i$ are independent eigenoperators, which as we proved above implies that they are degenerate. Thus, fermionic systems differ from spin systems in that they cannot be made to satisfy \eqref{eq:flucdissAB} without degeneracies in the spectrum of $\mathcal{L}$.

\textbf{Kramers' degeneracy in quantum channels}.~We briefly note that a similar Kramers' degeneracy can be found in thermal quantum channels, which can describe  the discrete time evolution of a system coupled to a \textit{non-Markovian} bath.  Define a quantum channel and its adjoint:
\begin{equation}
\hE(x) = \sum_i E_i x E_i^\dagger, \qquad \hE^\dagger(x) = \sum_i E_i^\dagger x E_i.
\end{equation}
The condition for trace preservation of the channel implies: $\hE^\dagger( \mathbb{I}) = \sum_i E_i^\dagger  E_i = \mathbb{I}$, which is the only condition that the Kraus operators ($E_i$) need to obey to be a proper channel. Suppose we further impose microreversibility: $ \hE^\dagger =   \hQ^{-1}  \hT ^{-1} \hE  \hT  \hQ$ where $\hQ,\hT$ are defined as before. It is easy to show that the thermal state $q$ is a steady state (eigenoperator of $\hE$ with eigenvalue $1$). For physical fermionic channels, the channel superoperator can be split into even and odd superparity sectors: $\hE = \text{Diag}[ \hE _+, \hE_- ]$. Our analysis implies that  the odd superparity sector $\hE_- $ must be twofold degenerate.

As an example,  the channel superoperator  could represent the  completely-positive-trace-preserving map for time-dependent Lindblad evolution: $\hE = \exp( \int \hL(t) dt )$. If the instantaneous Lindbladians $\hL(t)$ obey microreversibility at all times, then so should the channel superoperator $\hE$. The odd-superparity sector of  $\hE$ will then  have a twofold degeneracy.

\section{2.~Kramers' degeneracy in quadratic models}

We show that ``quadratic Lindbladians'' can host a twofold  degeneracy in their single-particle spectrum  in the absence of microreversibility, as long as  the system-environment coupling respects time-reversal symmetry. This is in contrast to the (quartic) models studied in the main text, where spectral splitting can emerge due to a non-equilibrium environment even if all couplings  respect TRS. The importance of microreversibility is therefore only apparent in quartic  models. 

Consider a quadratic Hamiltonian $H = \sum_{ij} H_{ij} \halo_i \halo_j $ in the presence of linear dissipators $L_\mu = \sum l_{\mu, i} \halo_i$, where $\halo_i$ are Majorana fermions. All terms in the master equation are quadratic in fermion operators, which implies that the Lindbladian  can be split into superparity sectors: $\hL = \hL_+ + \hL_-$. Define the superoperators: $e_j^\dagger (\rho ) = [\halo_j \rho  + (\hP \rho ) \halo_j]/2, e_j (\rho ) = [\halo_j \rho - (\hP \rho ) \halo_j]/2$,  where $\halo_j$ are Majoranas, and $\hP$ is the parity superoperator. Then we can express $\hL_{+}$ as
\begin{equation}
\hL_{+}  = 4 i \sum_{ij} (Z_{j i} e_i^\dagger e_j )+ (Y_{ij} e_i^\dagger e_j^\dagger )=   \sum_{i} \epsilon_i \hbetao^\dagger_i \hbetao'_i,
\end{equation}
where $Z=H + i \text{Re}[M]$, $Y=  \text{Im}[M]$, $M = l^T l^*$, $\epsilon_i/(4i)$ are the eigenvalues of  $Z$, and $\text{Re} [\epsilon_i] <0$ \cite{prosen2008, moos2019}. Analogously
\begin{equation}
\hL_{-}  = 4 i \sum_{ij} (Z_{j i} e_i e_j^\dagger ) + (Y_{ij} e_i e_j ) = \sum_{i} \epsilon_i +\sum_{i} (-\epsilon_i) \eta^\dagger_i \eta'_i. 
\end{equation}
The first term $\epsilon_m \equiv \sum_{i} \epsilon_i$ is a negative offset, then excitations have a \textit{positive} real energy.  The many-body eigenvalues are thus built from single-particle eigenvalues $\{ \epsilon_i \}$. 

Consider  a spin-$1/2$ Hamiltonian with TRS:
\begin{align}
H &= \frac{1}{2} (\mathbf{\halo})^T H \mathbf{\halo}, \qquad  \mathbf{\halo} = (a_{1,+}, b_{1,+}, \ldots, a_{1,-}, b_{1,-}  ,  \ldots)^T,
\end{align}
where $a,b$ are Majoranas which transform via $T a_{i,\sigma} T^{-1} =  \sigma a_{i,-\sigma}, T b_{i,\sigma} T^{-1} =  -\sigma b_{i,-\sigma} $, and $H = U H^* U^\dagger$. We include arbitrary linear dissipators which transform into each other (up to a phase) upon action of the symmetry operator:
\begin{equation}
L_{i,+}  = \sqrt{\gamma_i} f_{i, +}, \qquad L_{i,-}  = \sqrt{\gamma_i} f_{i,-}, \qquad T f_{i,+} T^{-1} = f_{i,-}, \qquad T f_{i,-} T^{-1} =- f_{i,+}.
\end{equation}
The dissipators can be expanded in terms of Majoranas: $L_{i,+} = \vec{l}_{i,+} \cdot \mathbf{\halo}, L_{i,-} = \vec{l}_{i,-} \cdot \mathbf{\halo}$,  which defines the matrix $M = l^T l^*$. We now show that $M = U M^* U^\dagger$:
\begin{equation}
U M^* U^\dagger = U (l^\dagger l) U^\dagger =    \left(\begin{array}{c}
 \vec{l}_{i,-} \\
 - \vec{l}_{i,+} \\
 \vdots
\end{array}\right)^T \left(\begin{array}{c}
 \vec{l}_{i,-} \\
 -\vec{l}_{i,+} \\
 \vdots
\end{array}\right)^*    =  M, 
\end{equation}
where we have used $U  \vec{l}_{i,+}^\dagger =  \vec{l}_{i,-}^T, U  \vec{l}_{i,-}^\dagger =  - \vec{l}_{i,+}^T$ in the second equality. In the last equality, we have used the fact that the Lindbladian is invariant under a relabeling and a change of sign of the dissipators.  This expression implies that the spectral matrix $Z=H + i \text{Re}[M]$ satisfies: $Z = U Z^T U^\dagger, U^2 = -\mathbb{I}$, which ensures that the single-particle spectrum is twofold degenerate.  Note that we have not  imposed microreversibility (detailed balance) for this result. Quadratic Lindbladians are thus special in the sense that the odd superparity sector can host a degeneracy in the absence of thermal equilibrium (unlike the examples in the main text).

\section{ 3.~Microreversibility from TRS-invariant system-bath coupling}

Here, we demonstrate that the Lindbladian of a Markovian open system will respect the microreversibility condition \eqref{eq:micro} if time-reversal symmetry is imposed on the system and bath as a whole. Before any Markovian approximation is made, the system and bath can be described by a Hamiltonian
\begin{align}
    H_{\rm tot} = H_S \otimes \mathbb{I}_B + \mathbb{I}_S \otimes H_B + H_{SB},
    \label{eq:hamTot}
\end{align}
where $H_S$, $H_B$ are the system and bath Hamiltonians, respectively, and $H_{SB}$ couples the two. Without loss of generality, we can decompose the latter as \cite{breuer2002}
\begin{align}
    H_{SB} = \sum_\alpha A_\alpha \otimes B_\alpha,
    \label{eq:HSBdecomp}
\end{align}
where $A_\alpha$, $B_\alpha$ are Hermitian matrices.

We suppose that the bath is initialized in a thermal Gibbs state $\rho_B = Z^{-1}_B e^{-\beta H_B}$, where $\beta = 1/T$ is the inverse temperature, and $Z_B = \Tr e^{-\beta H_B}$ is the partition function for the bath. Then define two-time correlators $\Gamma_{\alpha \beta}(t) = \Tr[B_\alpha(t) B_\beta(0) \rho_B]$, where
$B_\alpha(t) = e^{\iu H_B t}B_\alpha e^{-\iu H_B t}$. Hermiticity of $B_\alpha$ implies that
\begin{align}
    \Gamma_{\alpha \beta}(t)^* = \Gamma_{\beta \alpha}(-t).
    \label{eq:spectralHermiticity}
\end{align}
The expectation values $\braket{B_\alpha} \coloneqq \Tr B_\alpha \rho_B$ can always be made to vanish by replacing $B_\alpha \rightarrow B_\alpha - \braket{B_\alpha}$, and adding a term $A_\alpha$ to $H_S$, which does not change $H_{\rm tot}$. We also define the `lowering' operators $A_\alpha(\omega)$, which are the components of $A_\alpha$ that decrease the energy of the system by $\omega$. More concretely,
\begin{align}
    A_\alpha(\omega) = \sum_{\epsilon' - \epsilon = \omega} \Pi_\epsilon A_\alpha \Pi_{\epsilon'},
    \label{eq:AspectralDecomp}
\end{align}
where $\Pi_{\epsilon}$ is a projector onto the eigenspace of $H_S$ with eigenvalue $\epsilon$.

In order for the open system to be Markovian, the two-time correlation functions must decay over a timescale $\tau_{\rm m}$ that is sufficiently short, and the system-bath coupling $H_{SB}$ is sufficiently weak. If these criteria are met, then one can derive an expression for the Lindbladian in terms of the components of the microscopic Hamiltonian  \eqref{eq:hamTot} \cite{breuer2002}
\begin{align}
    \hL[\rho] = -\iu[H_S + H_{LS}, \rho] + \sum_{\omega} \sum_{\alpha \beta} \tilde{\Gamma}_{\alpha \beta}(\omega)\left( A_\beta(\omega) \rho A_\alpha(\omega)^\dagger - \frac{1}{2}\Big\{A_\alpha(\omega)^\dagger A_\beta(\omega),  \rho\Big\}\right),
\end{align}
where $\tilde{\Gamma}_{\alpha \beta}(\omega) = \int {\rm d} t e^{\iu \epsilon t} \Gamma_{\alpha \beta}(t)$ is the Fourier transform of the two-time correlation functions, which is a Hermitian matrix due to \eqref{eq:spectralHermiticity}. Here, we have defined the Lamb shift Hamiltonian $H_{LS} = \sum_{\omega, \alpha, \beta} S_{\alpha \beta}(\omega) A_\alpha(\omega) ^\dagger A_\beta(\omega)$, where $S_{\alpha \beta}(\omega) = \int {\rm d} t\, \text{sgn}(t) e^{\iu \epsilon t} \Gamma_{\alpha \beta}(t)$, which is Hermitian and commutes with $H_S$.

Now, TRS of the combined system and bath implies that $H_S$ and $H_B$ are each TRS-invariant, and that $\mathcal{T}_{SB}[H_{\rm tot}] = H_{\rm tot}$, where the superoperator $\mathcal{T}_{SB}$ acts as $\mathcal{T}_{SB}[O] = (U_S \otimes U_B) O^* (U_S^\dagger \otimes U_B^\dagger)$ for any operator $O$ over the system-bath Hilbert space. The operator $U_S$ is the unitary part of the TRS transformation acting on the system, as appears in Eq.~\eqref{eq:blockdiaghamiltonian}, and similarly $U_B$ is a unitary operator acting on the bath, which we leave unspecified for full generality. In terms of the decomposition \eqref{eq:HSBdecomp}, we have
\begin{align}
    U_S [A_\alpha]^* U_S^\dagger &= \sum_\beta u_{\alpha \beta} A_\beta & U_B [B_\alpha]^* U_B^\dagger &= \sum_\beta [u^{-1}]_{\beta \alpha} B_\beta
    \label{eq:TRScoupleOps}
\end{align}
for some matrix $u$. Hermiticity of $A_\alpha$, $B_\alpha$ implies that $u$ is a real matrix, and $\mathcal{T}^2 = \mathcal{P}$ implies that $[u u^*]_{\alpha \beta} = p_\alpha \delta_{\alpha \beta}$, where $p_\alpha = +1$ ($-1$) if $A_\alpha$ is a fermion parity even (odd) operator. The above conditions generalise the notion of `weak' symmetries described in Ref.~\cite{buca2012} to include antiunitary symmetry operations, with the difference that we impose restrictions on the microscopic Hamiltonian, rather than the emergent master equation.

Now, if the bath is in thermal equilibrium at inverse temperature $\beta$, then this imposes a Kubo-Martin-Schwinger (KMS) condition (or `detailed balance') on the spectral functions $\Gamma_{\alpha \beta}(\omega)$ \cite{kubo1957}:
\begin{align}
    \tilde{\Gamma}_{\alpha \beta}(-\omega) = e^{-\beta \omega}\tilde{\Gamma}_{\beta \alpha}(\omega).
    \label{eq:KMS}
\end{align}
Furthermore, since $H_B$ is TRS-invariant, we can use the transformation property \eqref{eq:TRScoupleOps} to determine how the two-time correlation functions transform under TRS
\begin{align}
    \Gamma_{\alpha \beta}(t) &= Z_B^{-1} \Tr\Big[ e^{\iu H_B t} B_\alpha e^{-\iu H_B t} B_\beta e^{-\beta H_B}\Big] \nonumber\\
    &= Z_B^{-1} \Tr\Big[U_B \left(e^{\iu H_B (t+\iu \beta)} B_\alpha e^{-\iu H_B t} B_\beta \right)^* U_B^\dagger\Big]^* \nonumber\\
    &= Z_B^{-1}\sum_{\gamma \delta} [u^{-1}]_{\gamma \alpha} [u^{-1}]_{\delta \beta}  \Tr\Big[ e^{-\iu H_B (t-\iu \beta)} B_\gamma e^{\iu H_B t} B_\delta  \Big]^* \nonumber\\
    &= Z_B^{-1}\sum_{\gamma \delta} [u^{-1}]_{\gamma \alpha} [u^{-1}]_{\delta \beta}  \Tr\Big[B_\delta e^{-\iu H_B t} B_\gamma  e^{\iu H_B (t+\iu \beta)}   \Big] \nonumber\\
    &= \sum_{\gamma \delta} [u^{-1}]_{\gamma \alpha} [u^{-1}]_{\delta \beta} \Gamma_{\delta \gamma}(t).
    \label{eq:spectralTRS}
\end{align}
The above can be Fourier transformed to obtain an analogous condition for $\tilde{\Gamma}_{\alpha \beta}(\omega)$.

With all these identities in hand, we are ready to begin our proof. We start by showing that the Gibbs state for the system $\rho_{G} = Z_S^{-1} e^{-\beta H_S}$ (where $Z_S = \Tr e^{-\beta H_S}$ is the system partition function) is a steady state. We have
\begin{align}
    \mathcal{L}[\rho_G] &= \frac{1}{Z_S} \sum_{\omega} \sum_{\alpha \beta} \tilde{\Gamma}_{\alpha \beta}(\omega)\left( A_\beta(\omega) e^{-\beta H_S} A_\alpha(\omega)^\dagger - \frac{1}{2}\Big\{A_\alpha(\omega)^\dagger A_\beta(\omega),  e^{-\beta H_S}\Big\} \right) \nonumber\\ 
    &= \frac{1}{Z_S} \sum_{\omega} \sum_{\alpha \beta} \tilde{\Gamma}_{\alpha \beta}(\omega) A_\beta(\omega) e^{-\beta H_S} A_\alpha(\omega)^\dagger - \frac{1}{2}\tilde{\Gamma}_{\beta \alpha}(\omega) \Big\{A_\beta(-\omega) A_\alpha(-\omega)^\dagger,  e^{-\beta H_S}\Big\} \nonumber\\
    &= \frac{1}{Z_S} \sum_{\omega} \sum_{\alpha \beta} \left(\tilde{\Gamma}_{\alpha \beta}(\omega) e^{-\beta \omega} A_\beta(\omega)  A_\alpha(\omega)^\dagger - \tilde{\Gamma}_{\beta \alpha}(\omega) A_\alpha(-\omega) A_\beta(-\omega)^\dagger   \right)e^{-\beta H_S} = 0.
\end{align}
In the first equality, we use $[H_{LS}, H_S] = 0$. The second equality involves a swap of labels $\alpha$ and $\beta$, and uses relation $A_\alpha(-\omega) = A_\alpha(\omega)^\dagger$, which can be verified using \eqref{eq:AspectralDecomp}. The third equality requires the relation $e^{-\beta H_S} A_\alpha(\omega)^\dagger = A_\alpha(\omega)^\dagger e^{-\beta (H_S+\omega)}$, and finally we replace $\omega \rightarrow -\omega$ in the second term and employ Eq.~\eqref{eq:KMS}.

Now we verify that Eq.~\eqref{eq:micro} is satisfied. Recall that $\mathcal{Q}[A] = q A$ for any operator $A$, and here $q = \rho_G$ is the steady-state density matrix. We can separate out $\hL = \hL_{\rm c} + \hL_{\rm d}$, where $\hL_{\rm c}[\rho] = -\iu [H_S + H_{LS}, \rho]$ contains the coherent part of the evolution, and the remainder $\hL_{\rm d}$ contains the dissipative part. Then the coherent part of the right-hand side of \eqref{eq:micro} acting on an arbitrary density operator $\rho$ gives (remembering that $\hT$ and $\hT^{-1}$ are antiunitary superoperators)
\begin{align}
    \hQ^{-1}\,  \hT ^{-1}\, \hL_{\rm c}\,  \hT\,  \hQ\,[\rho] &= \rho_G^{-1} \cdot  \hT ^{-1}\, \Bigg[-\iu\Big[H_S + H_{LS}, \hT[\rho_G \rho]\Big] \Bigg] \nonumber\\
    &=  +\iu \rho_G^{-1} \Big[\hT^{-1}[H_S + H_{LS}], \rho_G \rho\Big] = +\iu  \Big[H_S + H_{LS}, \rho_G^{-1} \rho_G \rho\Big] = \hL_{\rm c}^\dagger[\rho],
\end{align}
where we have used the fact that  that $H_S + H_{LS}$ is TRS-invariant and commutes with $\rho_G^{-1}$. Now the dissipative part is
\begin{align}
    \hQ^{-1}\,  \hT ^{-1}\, \hL_{\rm d}\,  \hT\,  \hQ\,[\rho] &=\rho_G^{-1} \cdot  \hT ^{-1}\, \Bigg[\sum_{\omega} \sum_{\alpha \beta} \tilde{\Gamma}_{\alpha \beta}(\omega) \left[ A_\beta(\omega) \rho_G \rho A_\alpha(\omega)^\dagger - \frac{1}{2}\Big\{A_\alpha(\omega)^\dagger A_\beta(\omega),  \hT[\rho_G \rho]\Big\}\right]\Bigg] \nonumber\\
    &= \rho_G^{-1} \sum_{\omega} \sum_{\alpha \beta \gamma \delta} [u^{-1}]_{\alpha \gamma} [u^{-1}]_{\beta \delta} \tilde{\Gamma}_{\alpha \beta}(\omega)^*\left( A_\delta(\omega) \rho_G \rho A_\gamma(\omega)^\dagger - \frac{1}{2}\Big\{A_\gamma(\omega)^\dagger A_\delta(\omega),  \rho_G \rho\Big\}\right) \nonumber\\
    &= \rho_G^{-1} \sum_{\omega} \sum_{\gamma \delta} \tilde{\Gamma}_{\gamma \delta}(\omega)\left( A_\delta(\omega) \rho_G \rho A_\gamma(\omega)^\dagger - \frac{1}{2}\Big\{A_\gamma(\omega)^\dagger A_\delta(\omega),  \rho_G \rho\Big\}\right) \nonumber\\
    &= \sum_{\omega} \sum_{\gamma \delta} \tilde{\Gamma}_{\gamma \delta}(\omega)\left( e^{-\beta \omega} A_\delta(\omega) \rho_G \rho A_\gamma(\omega)^\dagger - \frac{1}{2}\Big\{A_\gamma(\omega)^\dagger A_\delta(\omega),  \rho_G \rho\Big\}\right) \nonumber\\
    &= \sum_{\omega} \sum_{\gamma \delta} \tilde{\Gamma}_{\delta \gamma}(-\omega) A_\delta(\omega) \rho_G \rho A_\gamma(\omega)^\dagger - \frac{1}{2}\tilde{\Gamma}_{\gamma \delta}(\omega) \Big\{A_\gamma(\omega)^\dagger A_\delta(\omega),  \rho_G \rho\Big\},
    \label{eq:MicroDissLeft}
\end{align}
having used the transformation properties of $A_\alpha$ under $\hT^{-1}$ [Eq.~\eqref{eq:TRScoupleOps} with $u$ replaced by $u^{-1}$] in the second equality; Eq.~\eqref{eq:spectralTRS} in the third equality; and Eq.~\eqref{eq:KMS} in the final equality. This can be compared to $\hL_{\rm d}^\dagger$
\begin{align}
    \hL_{\rm d}^\dagger[\rho] &= \sum_{\omega} \sum_{\gamma \delta} \tilde{\Gamma}_{\gamma \delta}(\omega)^* \left( A_\delta(\omega)^\dagger \rho A_\gamma(\omega) - \frac{1}{2}\Big\{ A_\delta(\omega)^\dagger A_\gamma(\omega),  \rho\Big\}\right) \nonumber\\
    &= \sum_{\omega} \sum_{\gamma \delta} \tilde{\Gamma}_{\delta \gamma}(\omega) \left( A_\delta(-\omega) \rho A_\gamma(-\omega)^\dagger - \frac{1}{2}\Big\{ A_\delta(\omega)^\dagger A_\gamma(\omega),  \rho\Big\}\right),
    \label{eq:MicroDissRight}
\end{align}
having used $A_\alpha(-\omega) = A_\alpha(\omega)^\dagger$. Direct comparison verifies that \eqref{eq:MicroDissLeft} and \eqref{eq:MicroDissRight} are indeed equal, thus confirming that Eq.~\eqref{eq:micro} is satisfied.

\section{ 4.~Linear response in open quantum systems \label{secLinResp}}

This section shows that the retarded Green's function in Eq.~\eqref{eq:retardedgreens} in the main text  can be directly probed in  tunneling spectroscopy experiments (after taking the Fourier transform from the temporal to the frequency domain). 
We demonstrate that the standard formula for closed systems also applies to open quantum systems upon a generalization of operator time evolution. \\

\textbf{Kubo formula}. Consider a general Markovian dynamics in the form
\begin{align}\label{eq_supp:perturbed_lindblad}
\frac{d}{dt}\rho = \Bigl(\mathcal L_0+\lambda \mathcal L_1(t)\Bigl)\rho,
\end{align}
where $\mathcal L_0$ is the unperturbed Lindbladian, $\lambda\mathcal L_1(t)$ is time-dependent perturbation superoperator, and $\lambda$ is a small perturbation parameter.
Assume $\mathcal L_0(t) =\mathcal L_1\Theta(t-t_0)$ for some initial time $t_0$, where $\Theta(t-t')$ is the Heaviside step function. We also choose the initial state to be the steady state of the unperturbed system, $\rho(t_0) = \rho_{\rm SS}$ defined as $\mathcal L_0\rho_{\rm SS} = 0$.
In the lowest order of perturbative expansion over $\lambda$, the system's dynamics can be represented by a Dyson series,
\begin{align}\label{eq:dyson_expansion}
\begin{split}
\rho(t) &= \rho_{\rm SS} + \lambda\int_{t_0}^tdt'\exp\left(\mathcal L_0(t-t')\right)\mathcal L_1\exp(\mathcal L_0t')\rho_{\rm SS}+O(\lambda^2)\\
&= \rho_{\rm SS} + \lambda\int_{t_0}^tdt'\exp\left(\mathcal L_0(t-t')\right)\mathcal L_1\rho_{\rm SS}+O(\lambda^2).
\end{split}
\end{align}
Now let us consider the time-dependent expectation value of a local observable $ O$ defined as
$
O(t)= \Tr\left( O\rho(t)\right)
$.
Its dynamics can be expressed using Eq.~\eqref{eq:dyson_expansion} as
\begin{align}
\begin{split}
O(t) &= \langle  O\rangle_{\rm SS} + \lambda\int_{t_0}^tdt'\Tr\left(  O\exp\left(\mathcal L_0(t-t')\right)\mathcal L_1\rho_{\rm SS}\right)+O(\lambda^2)\\
&= \langle  O\rangle_{\rm SS} + \lambda\int_{t_0}^tdt'\Tr\left(\rho_{\rm SS}\mathcal L^\dag_1 \exp\left(\mathcal L^\dag_0(t-t')\right) O\right)+O(\lambda^2)\\
& = \langle  O\rangle_{\rm SS} + \lambda\int_{t_0}^tdt'\Tr\left(\rho_{\rm SS}\mathcal L^\dag_1  O(t-t')\right)+O(\lambda^2),
\end{split}
\end{align}
where $\langle  O\rangle_{\rm SS} = \Tr  O\rho_{\rm SS}$, $ O(t)=\exp(\mathcal L^\dag_0 t)  O$, and $\mathcal L^\dag_0,\mathcal L^\dag_1$ are conjugate Liouvillian operators as defined in Eq.~\eqref{eq:adjoint} in the main text.
We focus on unitary perturbation, 
$
\mathcal L_1 = -i[ H_1,\cdot\,]
$, where $ H_1$ is perturbation Hamiltonian. Extending $t_0\to-\infty$, we obtain
\begin{align}\label{eq_supp:kubo}
O(t) = \langle  O\rangle_{\rm SS} -i\lambda\int_{-\infty}^\infty dt'\Theta(t-t')\Bigl\langle[  O(t-t'), H_1]\Bigl\rangle_{\rm SS}+O(\lambda^2).
\end{align}
For unitary dynamics, this expression coincides with the conventional Kubo formula. \\

\textbf{Tunneling spectroscopy}. To measure the spectral properties of the system, we assume that the system of interest is connected to a tunneling probe as shown in Fig.~\ref{fig:bath} in the main text. The probe contains a reservoir of electrons able to tunnel into the system, thus generating an electric current. The dependence of the current on the probe's chemical potential, i.e. differential conductance, can be used to find the spectral function of the system.

Let us consider Eq.~\eqref{eq_supp:perturbed_lindblad} applied to a joint fermionic system-probe configuration. Without the perturbation, we assume that the system and the probe are decoupled, i.e. the unperturbed Liouvillian has the form
$
\mathcal L_0 = \mathcal L_0^S+\mathcal L_0^R
$,
where $\mathcal L_0^S$ and $\mathcal L_0^R$ are Liouville operators acting on system or the probe,   respectively. Also, we assume that the initial state is a product state, $\rho_{\rm SS} = \rho^S_{\rm SS}\otimes \rho^R_{\rm SS}$. The coupling between the system and the probe is produced by the perturbation 
\begin{align}
 H_1 = \sum_{\mu\nu}\bigl(T_{\mu\nu} f^\dag_\mu  b_\nu+{\rm h.c.}\bigl) =  B+ B^\dag,
\end{align}
where $ f_\mu$ and $ b_\nu$ are Fock operators for the system and the probe respectively, $T_{\mu\nu}$ are tunnelling coefficients, and $ B = \sum_{\mu\nu}T_{\mu\nu} f^\dag_\mu  b_\nu$. Here and below $\mu$ and $\nu$ are generalized indices that incorporate several quantum numbers, including the electron's position and spin.

The tunneling current is defined through the change of the probe's electron  number $N_D = \sum_\nu  b^\dag_\nu  b_\nu$, namely
\begin{align}
 I := i\lambda [ H_1,N_D] = -i\lambda\sum_{\mu\nu}\bigl(T_{\mu\nu} f^\dag_\mu  b_\nu-{\rm h.c.}\bigl) = -i\lambda( B- B^\dag).
\end{align}
Using the Kubo formula in Eq.~\eqref{eq_supp:kubo}, we obtain
\begin{align}\label{eq:current}
I(t) = 2\lambda^2{\rm Re}\int_{-\infty}^\infty dt'\Theta(t-t')\Bigl(\bigl\langle \bigl[ B^\dag(t-t'), B\bigl]\bigl\rangle_{\rm SS}-\bigl\langle \bigl[ B(t-t'), B\bigl]\bigl\rangle_{\rm SS}\Bigl),
\end{align}
where $ B(t) =\sum_{\mu\nu}T_{\mu\nu} f^\dag_\mu(t)  b_\nu(t)$ including $ f_\mu(t) := \eta_d \exp(\mathcal L^{S\dag}_0t)[\eta_d  f_\mu]$ and $ b_\nu(t) := \eta_d\exp(\mathcal L^{R\dag}_0 t)[\eta_d b_\nu]$, where $\eta_d$ is dummy Majorana fermion operator as defined in Eq.~\eqref{eqs:dummy_majorana} (see discussion in Section 1). Assuming that the probe is in the normal (i.e. not superconducting) state, the last term in Eq.~\eqref{eq:current} vanishes.
\begin{align}
\begin{split}
I(t) &= 2\lambda^2{\rm Re}\int_{-\infty}^\infty dt'\Theta(t-t')\sum_{\mu\nu}\sum_{\mu'\nu'}T^*_{\mu\nu}T_{\mu'\nu'}\Bigl(\langle  f^\dag_\mu(t-t') f_{\mu'}\rangle_{\rm SS}\langle  b_\nu(t-t') b^\dag_{\mu'}\rangle_{\rm SS}-\langle  f_\mu(t-t') f^\dag_{\mu'}\rangle_{\rm SS}\langle  b^\dag_\nu(t-t') b_{\mu'}\rangle_{\rm SS}\Bigl)\\
 &=2\lambda^2{\rm Re}\int_{-\infty}^0 d\tau \sum_{\mu\nu}\sum_{\mu'\nu'} T^*_{\mu\nu}T_{\mu'\nu'}\Bigl(G^<_{\mu\mu'}(\tau)D^>_{\nu\nu'}(-\tau)-G^>_{\mu\mu'}(\tau)D^<_{\nu\nu'}(-\tau)\Bigl)= \int_{-\infty}^\infty \frac{d\omega}{2\pi} I(\omega),
 \end{split}
\end{align}
where $G^<_{\mu\mu'}(\tau) = -i\langle  f^\dag_\mu(\tau) f_{\mu'}\rangle_{\rm SS}$ and $G^>_{\mu\mu'}(\tau) = -i\langle  f^\dag_\mu  f_{\mu'}(\tau)\rangle_{\rm SS}$ are the greater and the lesser Green's function, $D^<_{\nu\nu'}(\tau)$ and $D^<_{\nu\nu'}(\tau)$ are similar expressions for the probe in terms of $ b_\nu$, and $I(\omega)$ are the Fourier components of the tunneling current defined as
\begin{align}
I(\omega)=\lambda^2\sum_{\mu\nu}\sum_{\mu'\nu'} T^*_{\mu\nu}T_{\mu'\nu'} \Bigl(G^<_{\mu\mu'}(\omega)D^>_{\nu}(\omega)-G^>_{\mu\mu'}(\omega)D^<_{\nu}(\omega)\Bigl).
\end{align}
Let us assume that the probe's reservoir of fermions is at thermal equilibrium at zero temperature and $ b_\nu$ are eigenmodes of the reservoir. Then, the expression for probe's Green's functions takes the form
\begin{align}
D^>_{\nu\nu'}(\omega) =-i \delta_{\nu\nu'}\tilde A_\nu(\mu+\omega)\Bigl(1-n_F(\mu+\omega)\Bigl), \qquad D^<_{\nu\nu'}(\omega) = i \delta_{\nu\nu'} \tilde A_\nu(\mu+\omega)n_F(\mu+\omega),
\end{align}
where $n_F(x) = \Theta(x)$ is zero-temperature Fermi distribution, $\mu$ is probe's chemical potential, and $\tilde A_\nu$ is the probe's spectral function.

Assuming $\tilde A_\nu(V+\omega) \simeq \tilde A_\nu$ depends weakly on the chemical potential $V$, and introducing coefficients $J_{\mu\mu'} = 2\lambda^2 \sum_{\nu} T^*_{\mu\nu}T_{\mu'\nu} \tilde A_{\nu}$, we obtain the expression for the differential conductance
\begin{align}
\frac{\partial I(\omega)}{\partial \mu}= \sum_{\mu\mu'} J_{\mu\mu'}{\rm Im}\, G^R_{\mu\mu'}(\omega)\delta(\omega+\mu),
\end{align}
where $\delta(x)$ is the Dirac delta function, $G^R_{\mu\mu'}(\omega) = G^<_{\mu\mu'}(\omega)+G^>_{\mu\mu'}(\omega)$ is the Fourier transform of the time-domain retarded Green's function defined in Eq.~\eqref{eq:retardedgreens} in the main text. Choosing appropriate filtering $J_{i\sigma',j\sigma''}\propto \delta_{ij}\delta_{\sigma'\sigma''}\delta_{\sigma \sigma'}$, we can measure spin-polarized current characteristics,
\begin{align}
\frac{\partial I_\sigma}{\partial \mu} \propto {\rm Im}\, G^R_{i\sigma}(-\mu).
\end{align}
Thus, changing the chemical potential $\mu$, we can probe the spectral function of the system.

\end{document}